\documentclass[journal]{IEEEtran}
\usepackage{graphicx}
\usepackage{subfigure}
\usepackage{float}
\usepackage{amsmath}
\usepackage{epstopdf}
\usepackage{cases}
\usepackage{color}
\usepackage{algorithm}
\usepackage{algpseudocode}
\usepackage{amsmath}
\usepackage{amssymb}

\makeatletter
\renewcommand{\maketag@@@}[1]{\hbox{\m@th\normalsize\normalfont#1}}%
\makeatother
\usepackage{stfloats}
\usepackage{cite}
\usepackage{makecell}
\usepackage{multirow}

\ifCLASSINFOpdf
\else
\fi

\usepackage{caption}
\usepackage{mathtools}

\begin{document}

\title{Integrated Sensing and Communication Signals Toward 5G-A and 6G: A Survey}

\author{
Zhiqing~Wei,~\IEEEmembership{Member,~IEEE,}
Hanyang~Qu,~\IEEEmembership{Member,~IEEE,}
Yuan~Wang,~\IEEEmembership{Member,~IEEE,}
Xin~Yuan,~\IEEEmembership{Member,~IEEE,}
Huici~Wu,~\IEEEmembership{Member,~IEEE,}
Ying~Du,~\IEEEmembership{Member,~IEEE,}
Kaifeng~Han,~\IEEEmembership{Member,~IEEE,}\\
Ning~Zhang,~\IEEEmembership{Senior,~IEEE,}
Zhiyong~Feng,~\IEEEmembership{Senior,~IEEE}

\thanks{This work was supported in part by the National Natural Science Foundation of China (NSFC) under Grant 62271081, 92267202 and U21B2014, in part by the National Key Research and Development Program of China under Grant 2020YFA0711302, and in part by the Young Elite Scientists Sponsorship Program by CAST under Grant YESS20200283.

Zhiqing Wei, Hanyang Qu, Yuan Wang, and Zhiyong Feng are with the Key Laboratory of Universal Wireless Communications, Ministry of Education, School of Information and Communication Engineering, Beijing University of Posts and Telecommunications, Beijing 100876, China (emails: \{weizhiqing; hanyangqu; wangyuan; fengzy\}@bupt.edu.cn).

Huici Wu is with the National Engineering Lab for Mobile Network Technologies, Beijing University of Posts and Telecommunications, Beijing 100876, China, and also with Peng Cheng Laboratory, Shenzhen 518066, China (e-mail: dailywu@bupt.edu.cn).

Xin Yuan is with Data61, CSIRO, Sydney, Australia (email: xin.yuan@data61.csiro.au).

Ning Zhang is with the Department of Electrical and Computer Engineering, University of Windsor, Windsor, ON, N9B 3P4,
Canada. (email: ning.zhang@uwindsor.ca).

Ying Du is with the University of Science and Technology of China,
and China Academy of Information and Communications Technology, Beijing 100191, China.
Kaifeng Han is with China Academy of Information and Communications Technology,
Beijing 100191, China.
(emails: \{duying1; hankaifeng\}@caict.ac.cn).

Correspondence authors: Ying Du, Zhiyong Feng, Zhiqing Wei.}}

\maketitle

\begin{abstract}
Integrated sensing and communication (ISAC) has the advantages of efficient
spectrum utilization and low hardware cost.
It is promising to be implemented in the
fifth-generation-advanced (5G-A) and sixth-generation (6G) mobile communication systems,
having the potential to be applied in intelligent applications requiring both communication
and high-accurate sensing capabilities.
As the fundamental technology of ISAC, ISAC signal directly impacts the performance of
sensing and communication.
This article systematically reviews the literature on ISAC signals from the perspective of
mobile communication systems, including ISAC signal design,
ISAC signal processing and ISAC signal optimization.
We first review the ISAC signal design based on 5G, 5G-A and 6G mobile communication systems.
Then, radar signal processing methods are reviewed for ISAC signals,
mainly including the channel information matrix method,
spectrum lines estimator method and super resolution method.
In terms of
signal optimization, we summarize peak-to-average power ratio (PAPR) optimization,
interference management, and adaptive
signal optimization for ISAC signals. This article may provide the guidelines for
the research of ISAC signals in 5G-A and 6G
mobile communication systems.
\end{abstract}


\IEEEpeerreviewmaketitle

\begin{figure*}[!htbp]
	\centering
	\includegraphics[width=1\textwidth]{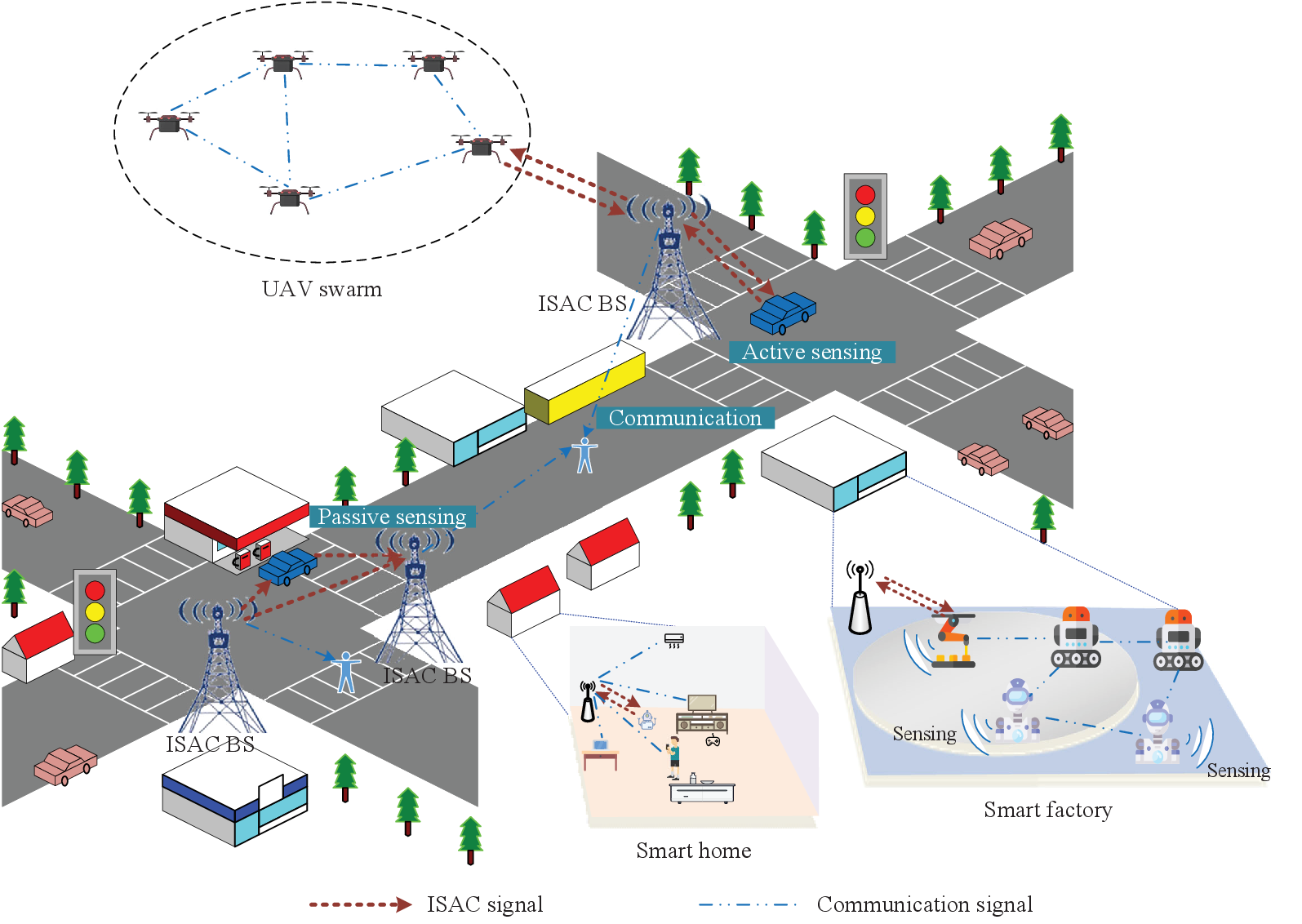}
	\caption{Application scenarios of ISAC in the future.}
	\label{Rich ISAC scenarios}
\end{figure*}

\begin{table*}[!htbp]
	\caption{Glossary}
	\label{tab_g}
	\renewcommand{\arraystretch}{1.5} 
	\begin{center}
		\begin{tabular}{m{0.1\textwidth} m{0.3\textwidth} m{0.1\textwidth} m{0.3\textwidth}}
			\hline
			\hline
			5G-A & Fifth-generation-Advanced &
			6G & Sixth-generation\\
			APAR & Active Phased Array Radar &
			AWGN & Additive White Gaussian Noise \\
			BS & Base Station &
			BER & Bit Error Rate \\
			CP & Cyclic Prefix &
			CC & Cyclic Cross-correlation \\
			CD-OFDM & Code Division OFDM &
			CSMA & Carrier Sense Multiple Access \\
			CSMA/CA & Carrier Aware Multiple Access / Collision Avoidance &
			CSI & Channel State Information \\
			DSSS & Direct Sequence Spread Spectrum &
			DARPA & Defense Advanced Research Projects Agency \\
			DFRC & Dual-functional Radar-Communication &
			DoA & Directions of Arrival \\
			DAS & Distributed Antenna System &
			DIR & Data Information Rate \\
			ESPRIT & Estimating Signal Parameter via Rotational Invariance Techniques &
			EM & Expectation Maximization \\
			FBMC & Filter-bank Multi-carrier &
			FM & Frequency Modulation \\
			FRFT & Fractional Fourier Transform &
			FCW & Forward Collision Warning \\
			GFDM & Generalized Frequency Division Multiplexing &
			ICI & Inter-carrier Interference \\
			IoV & Internet of Vehicle &
			ISAC & Integrated Sensing and Communication  \\
			ISI & Inter-symbol Interference &
			KKT & Karush Kuhn Tucher \\
			LRR & Long-Range Radar &
			MTC & Machine-Type Communication \\
			mmWave & Millimeter-wave &
			MIMO & Multiple Input Multiple Output \\
			MUSIC & Multiple Signal Classification &
			MTI & Moving Target Indication \\
			MSK & Minimum-Shift Keying &
			ML & Maximum Likelihood \\
			MI & Mutual Information &
			MMSE & Minimum Mean Square Error \\
			MRR & Medium-range Radar &
			MVDR & Minimum Variance Distortionless Response \\
			MC-DS-CDMA & Multicarrier Direct Sequence CDMA &
			NASA & National Aeronautics and Space Administration \\
			NC-OFDM & Non-Contiguous-Orthogonal Frequency-Domain Modulation &
			OFDM & Orthogonal Frequency Division Multiplex \\
			OTFS & Orthogonal Time Frequency Space &
			OOB & Out of Band \\
			OCDM & Orthogonal Chirp Division Multiplexing &
			OFDMA & Orthogonal Frequency Division Multiple Access \\
			PAPR & Peak to Average Power Ratio &
			PSLR & Peak Side Lobe Ratio \\
			PMCW & Phase Modulated Continuous Waveform &
			PTS & Partial Transmission Sequence \\
			QCQFP & Quadratically Constrained Quadratic Fractional Programming &
			RCC & Radar-Communication Coexistence \\
			SSPARC & Shared Spectrum Access for Radar and Communications &
			SIR & Signal to Interference Ratio \\
			SS-OFDM & Spread Spectrum OFDM &
			SIC & Serial Interference Cancellation \\
			SLM & Selected Mapping &
			S\&C & Schmidl and Cox \\
			SCM & Slow Chirp Modulation &
			SPR & Subcarrier Power Ratio \\
			SRR & Short-range Radar &
      		SVD & Singular Value Decomposition \\
			TLS-SVD & Total Least Squares-Singular Value Decomposition &
			THz & Terahertz \\
			V2V & Vehicle-to-Vehicle &
			XR & Extended Reality \\
			\hline
			\hline
		\end{tabular}
	\end{center}
\end{table*}

\section{Introduction}

In the future 5G-A and 6G mobile communication systems,
new intelligent applications and services have emerged,
such as machine-type communication (MTC), connected robotics and autonomous driving,
and extended reality (XR) \cite{feng2020joint}.
Currently, separated design of sensing and communication cannot achieve high data transmission
and high-accurate sensing simultaneously and meet the requirements of
these new emerging services and applications \cite{csahin2020multi}.
Besides, with increasingly scarce spectrum resources,
it is difficult to satisfy the spectrum requirements of the
new intelligent applications and services.

Integrated sensing and communication (ISAC) system is a unified system providing
wireless communication and radar sensing functions.
ISAC effectively improves the spectrum utilization and solves the spectrum conflict
between radar and communication systems. In addition,
ISAC system reduces the size and energy consumption of the equipment.
Due to the high spectrum efficiency and low hardware cost, academia and
industry
both pay extensive attention to ISAC technology \cite{hayvaci2014spectrum,chiriyath2017radar,liu2020joint,zhang2021overview}.
As shown in Fig. \ref{Rich ISAC scenarios},
ISAC technology is expected to be applied in a wide range of
intelligent applications that require high communication rates and
high-accurate sensing capabilities in the era of 5G-A and 6G.
With the development and application of millimeter-wave (mmWave) and terahertz (THz)
technologies, the frequency bands of communication gradually coincide with these of radar,
which promotes the realization of ISAC technology \cite{griffiths2014radar}.

ISAC technology is developed on the basis of active phased array radar (APAR), since APAR is easily integrated with communication technologies \cite{quan2014radar}.
In 1963, Mealey \emph{et al.} proposed that radar pulses can transmit
information to space vehicles \cite{mealey1963method}.
In the 1980s, the national aeronautics and space administration (NASA) begin
to study the ISAC technology to provide communication and navigation for the space shuttle.
A space shuttle program is proposed to support the research of ISAC \cite{scharrenbroich2016joint}.
Since the 1990s, multiple countries have competed to carry out the research of ISAC technology.
The US Office of Naval Research launches the advanced multi-function radio frequency concept program in 1996, trying to integrate radar,
electronic warfare, communication and other functions by using the broadband radio frequency front-end aperture with separated transmitter (TX) and receiver (RX).
Germany and Britain have also started the research on ISAC.
In 5G-A and 6G, the ISAC enables mobile communication systems have attracted
much attention \cite{barneto2019full}.
international mobile telecommunication (IMT) 2030 \cite{IMT}
has regarded ISAC as one of the potential technologies of 6G.

ISAC signals, as the basis of ISAC technology, have been studied extensively.
In 2006, Shao \emph{et al.} apply direct sequence spread spectrum (DSSS) technique to spread the spectrum of the ISAC system to avoid the jamming between radar
and communication \cite{shaojian2006radar}.
In 2011, Sturm \emph{et al.} design an ISAC signal based on
orthogonal frequency division multiplex (OFDM) \cite{sturm2009ofdm}.
They demonstrate that OFDM has a high dynamic range of
radar imaging \cite{sturm2010performance}, which is able to detect multiple targets \cite{sit2011extension}.
In 2013, the shared spectrum access for radar and communications (SSPARC)
project proposed by defense advanced research projects agency (DARPA)
promotes the development of ISAC technology \cite{zhou2018resource}.
In 2014, Koslowski \emph{et al.} design a filter-bank multicarrier (FBMC) based ISAC
signal, and show that the sensing performance of FBMC is similar to that of OFDM.
FBMC-based ISAC signal reduces the overhead of cyclic prefix (CP) and
improves the accuracy of velocity estimation \cite{koslowski2014using}.
In 2017, the ISAC signal based on orthogonal time frequency space (OTFS) is designed to
detect the target with high mobility \cite{hadani2017orthogonal}.
In 2019, Sanson \emph{et al.} design generalized frequency division multiplexing (GFDM)
based ISAC signal, where the signal can reduce out of band (OOB) and
use the spectrum without interfering with other users in
comparison with OFDM \cite{sanson2019non}.

Nevertheless, the design of ISAC signal faces great challenges, which are summarized as follows.

\begin{itemize}
	\item[$\bullet$] \textbf{ISAC signal design}: Communication aims to 
	achieve efficient and reliable data transmission, 
	which requires high spectrum efficiency and the capability of 
	anti-interference and combating channel fading. 
	Radar is to achieve high-resolution target sensing, 
	which requires good autocorrelation, large signal bandwidth, large dynamic range, 
	and large Doppler frequency shift. Thus, the ISAC signal design needs to balance 
	the performance of radar and communication \cite{yuan2020spatio,yuan2020Waveform}.
\end{itemize}
\begin{itemize}
	\item[$\bullet$] \textbf{ISAC signal processing}: In order to improve the sensing accuracy 
	with limited computation resources, 
	the design of ISAC signal processing algorithms with 
	high-accuracy and low-complexity is required.
\end{itemize}
\begin{itemize}
	\item[$\bullet$] \textbf{ISAC signal optimization}: It is necessary to design flexible ISAC signal optimization to meet the requirements of sensing and communication in various scenarios.
\end{itemize}

Facing the above challenges, some existing articles have reviewed the ISAC technology.
Hayvaci \emph{et al.} review the ISAC signals from the perspectives of cognitive communication, cognitive radar, and joint cognition of radar and communication systems \cite{hayvaci2014spectrum}.
Chiriyath \emph{et al.} define three levels
of ISAC, including coexistence, cooperation, and co-design.
In addition, the interference management methods between
radar and communication are summarized \cite{chiriyath2017radar}.
Liu \emph{et al.} comprehensively review the scenarios and key technologies of
radar communication coexistence (RCC) and dual function radar communication (DFRC) \cite{liu2020joint}.
Zhang \emph{et al.} summarize radar signal processing algorithms in ISAC system \cite{zhang2021overview}.
\cite{zhang2021enabling} reviews the development status
and challenges of ISAC in the context of perceptive mobile networks.
\cite{wang2021symbiotic} analyzes the key technologies of ISAC signal in 6G from different application requirements.
\cite{tan2021integrated} proposes four typical use cases of ISAC,
where the requirements of these use cases are analyzed in detail.
The existing ISAC signals are reviewed from the aspect of
signal design \cite{xiong2022overview}.
The application scenarios and
key technologies of ISAC are summarized in \cite{zhang2021enabling,liu2020joint}.
Recently, institute of electrical and electronics engineers (IEEE),
3rd generation partnership project (3GPP), and IMT-2030, etc., have
launched projects on ISAC technology.

The above studies provide a comprehensive overview on the scenarios,
requirements and key technologies of ISAC. However, the ISAC signals from the perspective
of 5G-A and 6G mobile communication systems have not been thoroughly reviewed.
As 6G takes ISAC as the potential key technology and
the ISAC signals are
the foundation of ISAC technology, the ISAC signals from the perspective of
mobile communication systems of 5G-A and 6G have attracted extensive attention.
In this article, we review the ISAC signals, including ISAC signal design,
ISAC signal processing, and ISAC signal optimization.
The contributions of this article are summarized as follows.
\begin{itemize}
	\item[$\bullet$] \textbf{ISAC signal design}: This article summarizes ISAC signal design based on the signals in 5G, 5G-A and 6G mobile communication systems, such as the ISAC signal based on OFDM \cite{quan2014radar,6986232}, FBMC \cite{koslowski2014using,cao2016feasibility}, GFDM \cite{sanson2019non}, discrete fourier transform-spread-orthogonal frequency division multiplexing (DFT-s-OFDM) \cite{gerzaguet20175g}, and OTFS \cite{raviteja2018interference,gaudio2020effectiveness}.
\end{itemize}
\begin{itemize}
	\item[$\bullet$] \textbf{ISAC signal processing}: The signal processing algorithms for ISAC signal are introduced, including channel information matrix method, spectrum lines estimator method and super resolution method. Then, the detailed signal processing algorithms for the OFDM and OTFS-based ISAC signals are reviewed.
\end{itemize}
\begin{itemize}
	\item[$\bullet$] \textbf{ISAC signal optimization}: The ISAC signal optimization methods, including PAPR reduction, interference management and adaptive signal optimization, are reviewed.
\end{itemize}

\begin{figure}[!h]
	\centering
	\includegraphics[width=0.5\textwidth]{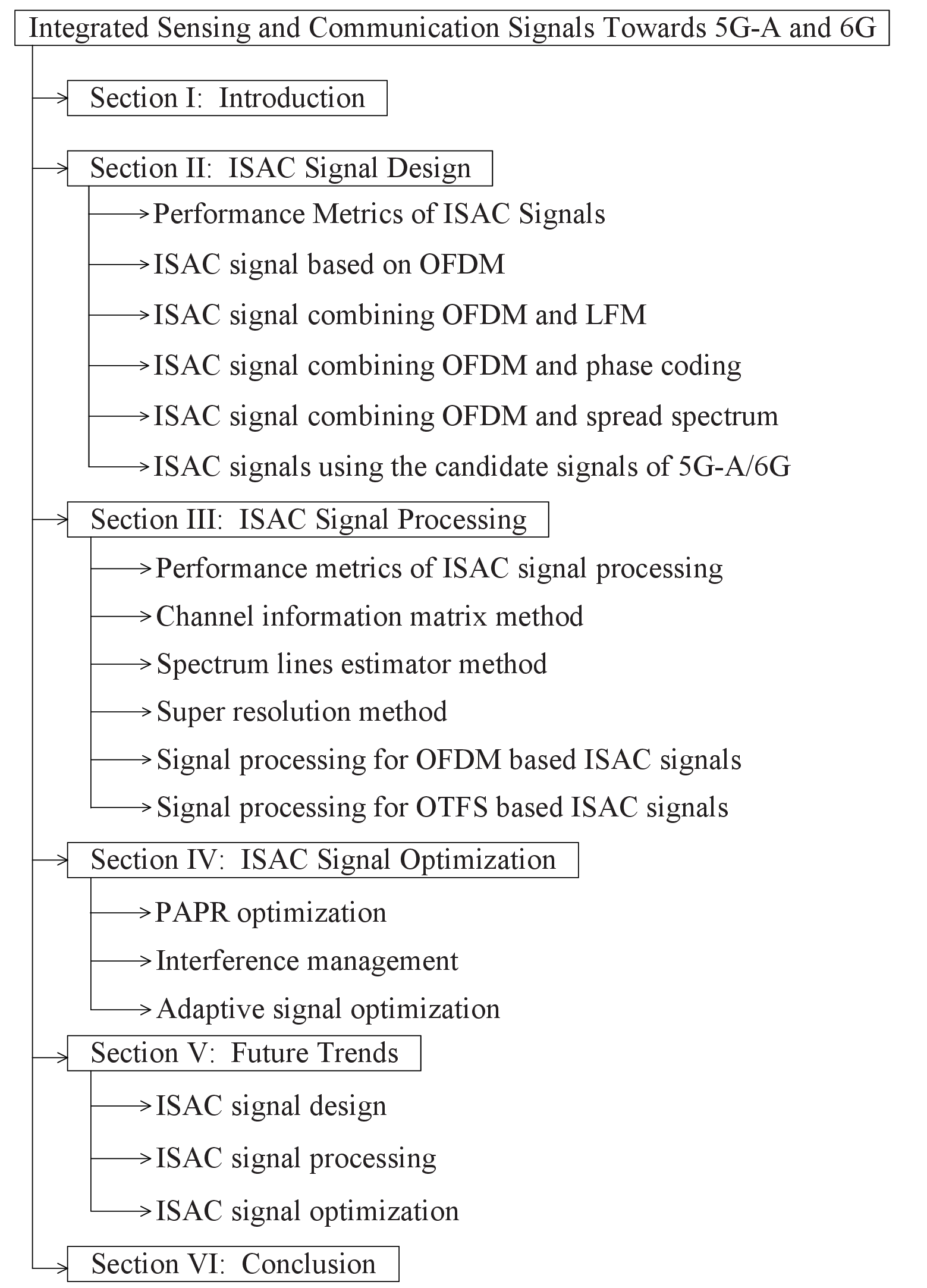}
	\caption{The organization of this article.}
	\label{The organization of this article}
\end{figure}
The organization of this article is shown in Fig. \ref{The organization of this article}.
In Section \uppercase\expandafter{\romannumeral2}, we review ISAC signal design based on mobile communication systems.
In Section \uppercase\expandafter{\romannumeral3}, we review the ISAC signal processing methods.
The ISAC signal optimization methods are summarized in Section \uppercase\expandafter{\romannumeral4}.
The future trends are provided in Section \uppercase\expandafter{\romannumeral5}, followed by conclusions in Section \uppercase\expandafter{\romannumeral6}.
The glossary of this article is provided in Table \ref{tab_g}.

\section{ISAC Signal Design}\label{sec:system-model}
OFDM is the signal of the 4th-generation (4G) and 5G mobile communication systems. 
The candidate signals of the next-generation mobile communication systems include FBMC, GFDM, 
DFT-s-OFDM, OTFS, etc. Various ISAC signal design schemes are proposed 
using the above-mentioned signals. Firstly, we introduce the performance metrics of ISAC signals, 
followed by the features of OFDM. Then, ISAC signals based on OFDM are introduced. 
Finally, ISAC signals based on the candidate signals of the next-generation mobile 
communication system are reviewed.

\subsection{Performance Metrics of ISAC Signals}
\begin{figure*}[!htbp]
	\centering
	\includegraphics[width=0.99\textwidth]{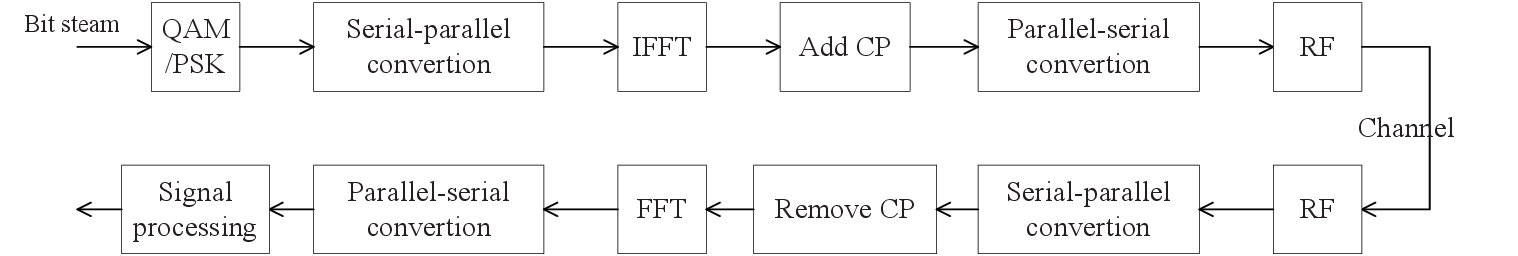}
	\caption{Signal processing procedure for OFDM signal \cite{quan2014radar}.}
	\label{Signal processing procedures for OFDM signals}
\end{figure*}
This section summarizes the performance metrics of ISAC signals, which are used to evaluate the performance of ISAC signals for various application scenarios.

\begin{itemize}
	\item[$\bullet$] \textbf{Resolution}: Resolution reveals the ability of ISAC signal to distinguish multiple targets. A small value of resolution of the ISAC signal indicates that the signal has a strong ability to distinguish multiple targets \cite{wehner1987high}.
\end{itemize}

\begin{itemize}
	\item[$\bullet$] \textbf{Ambiguity function}: Ambiguity function is the time-frequency composite autocorrelation function of the complex envelope of ISAC signal. Under the conditions of optimal signal processing, the performance metrics of resolution, ambiguity, accuracy and clutter suppression of radar sensing are derived via ambiguity function~\cite{stein1981algorithms}.
\end{itemize}

\begin{itemize}
	\item[$\bullet$] \textbf{Doppler sensitivity}: Doppler sensitivity is the error of output response of the radar signal processing module and communication demodulation module caused by the Doppler frequency shift.
	For the signals with high Doppler sensitivity, the sensing output of the radar signal processing module has an intolerable error in the presence of a Doppler frequency shift.
	Besides, the Doppler frequency shift affects the bit error rate (BER) of the received signals \cite{zhao1996sensitivity,johnston1986waveform}.
\end{itemize}

\begin{itemize}
	\item[$\bullet$] \textbf{Doppler tolerance}: Doppler tolerance is obtained by calculating the cross-correlation between stretched echoes and narrowband approximation signal subtraction in time domain. Doppler tolerance reveals the maximum velocity detected by the radar, beyond which the velocity estimation will produce intolerable error \cite{franken2006doppler}.
\end{itemize}

\begin{itemize}
	\item[$\bullet$] \textbf{PAPR}: PAPR is the ratio of the peak instantaneous
power to the average power. A high PAPR will require a
high-performance power amplifier, resulting in additional
overhead for hardware equipment \cite{han2004papr}.
\end{itemize}

\begin{itemize}
	\item[$\bullet$] \textbf{Mutual information (MI)}: Communication MI evaluates
the channel capacity. Sensing MI is a performance metric that measures the performance of ISAC signals. The
sensing MI is defined by the conditional MI of the
sensing channel and the echo signal, which is used as
the evaluation criterion of sensing performance \cite{yuan2020spatio}.
\end{itemize}

\begin{itemize}
	\item[$\bullet$] \textbf{Data information rate (DIR)}: DIR is the information transmission rate, which is used to evaluate the communication efficiency of ISAC signal \cite{liu2017adaptive}.
\end{itemize}

\subsection{ISAC Signal based on OFDM}
\begin{table*}[htbp]
	\caption{ Characteristics of OFDM-based ISAC signals (\checkmark indicates the existence of the advantage)}
	\renewcommand{\arraystretch}{1.7}
	\begin{center}
		\begin{tabular}{|m{0.12\textwidth}<{\centering}|m{0.24\textwidth}|m{0.13\textwidth}|m{0.11\textwidth}|m{0.11\textwidth}|m{0.06\textwidth}|m{0.06\textwidth}|}
			\hline
			\textbf{OFDM based ISAC signal} & \makecell[c]{\textbf{Design method}} & \makecell[c]{\textbf{Resolution}} & \makecell[c]{\textbf{PAPR reduction}} & \makecell[c]{\textbf{Anti-interference}} & \makecell[c]{\textbf{Secrecy}} & \makecell[c]{\textbf{References}}\\
			\hline
			Combination of OFDM and LFM             & LFM increases the time-bandwidth product of the signal. & \makecell[c]{\checkmark}         &             &              &    &\makecell[c]{\cite{li2018communication,alabd2019partial,zhang2017rf,striano2019performance}}        \\
			\hline
			Combination of OFDM and phase coding    & Phase coding provides a flexible signal design to reduce PAPR. & \makecell[c]{\checkmark} & \makecell[c]{\checkmark}  &  &  &\makecell[c]{\cite{tian2017waveform,zhao2014chaos,dokhanchi2018multicarrier,uysal2019phase}}         \\
			\hline
			Combination of OFDM and spread spectrum & Spread spectrum improves anti-jamming performance. &                  &             & \makecell[c]{\checkmark}   & \makecell[c]{\checkmark} & \makecell[c]{\cite{nusenu2018time,kim2012novel,cheng2015spread,sharma2020multicarrier}} \\
			\hline
		\end{tabular}
	\end{center}
\end{table*}

\begin{table*}[htbp]
	\caption{ OFDM-LFM signals}
	\renewcommand{\arraystretch}{1.7}
	\begin{center}
		\begin{tabular}{|m{0.22\textwidth}<{\centering}|m{0.15\textwidth}|m{0.55\textwidth}|}
			\hline
			\textbf{Category} & \makecell[c]{\textbf{References}} & \makecell[c]{\textbf{One sentence summary}}\\
			\hline
			Carrier modulation & \makecell[c]{\cite{sidney1954pulse,4101252}} & The subcarrier of OFDM is designed in the form of LFM, and the communication symbol is modulated to the subcarrier.\\
			\hline
			
			\multirow{2}[4]{\linewidth}{\makecell[c]{Non-carrier modulation}}
			& \makecell[c]{\cite{takahashi2019novel}} & The communication signal is multiplexed at the sideband of an LFM pulse to improve the communication rate. \\
			\cline{2-3}
			& \makecell[c]{\cite{dash2015generalized}} & A generalized OFDM-LFM signal is designed to obtain a large time-bandwidth product. \\
			\hline
		\end{tabular}
	\end{center}
\end{table*}
OFDM is a kind of multi-carrier modulation technology, whose signal processing procedure is shown in Fig. \ref{Signal processing procedures for OFDM signals}.
The ISAC signal is formulated as \cite{quan2014radar}

\begin{equation}
	\begin{aligned}
		&s\left( t \right) = \sum\limits_{m = 0}^{M - 1} {\sum\limits_{n = 0}^{N - 1} } {\Huge (} {d\left( {m N + n} \right)}  \\
		&\quad \quad \quad  \cdot \exp \left( {j2\pi {f_n}t} \right){\rm{rect}}( {\frac{{t - m {T_{sym}}}}{{{T_{sym}}}}} ){\Huge )},
	\end{aligned}
\end{equation}
where ${M}$ is the number of symbols, ${N}$ is the number of subcarriers, ${T_{sym}}$ is the OFDM symbol duration, ${f_n}$ is the frequency of the $n$-th subcarrier, $d\left( {m N + n} \right)$ is the modulation symbol, and ${\rm{rect}}\left(  \cdot  \right)$ is rectangular window function.

The subcarriers of OFDM are orthogonal to each other and have the same frequency spacing. 
The orthogonality reduces inter-carrier interference (ICI), 
and improves the spectrum efficiency. 
OFDM facilitates synchronization and equalization, 
because of the capability of combatting multipath fading. 
The subcarriers in the ISAC signal using OFDM are orthogonal, 
and the number and frequency spacing of the subcarriers are flexibly adjusted according 
to the requirements of various scenarios. As for the sensing capability, 
a thumbtack-typed ambiguity function is obtained to reduce 
the delay-Doppler ambiguity for moving target detection, and high-resolution imaging \cite{6986232}. 
Therefore, OFDM is widely applied in the design of ISAC signals due to its 
excellent communication and sensing performance.

However, the OFDM signal has high PAPR, distorting the RF front-end \cite{quan2014radar}.
Meanwhile, the mutual-interference between the radar echo signal and communication signal, as well as the self-interference between TX and RX, both degrade the performance of  radar sensing and communication in the ISAC system, which needs to be solved via signal optimization.

OFDM is usually combined with other modulation methods such as linear frequency modulation (LFM), phase coding, and spread spectrum to improve the sensing performance of OFDM-based ISAC signals.
The OFDM signals combined with these modulation methods have better sensing performance improvement in terms of optimizing ambiguity function, improving anti-interference capability, and reducing PAPR.
The ISAC signal combining OFDM and LFM is proposed to improve the resolution of detecting long-distance targets, and the resolution is improved by increasing the time-bandwidth product \cite{li2018communication,alabd2019partial,zhang2017rf,striano2019performance}.
ISAC signal combining OFDM and phase coding reduces the PAPR \cite{tian2017waveform,zhao2014chaos,dokhanchi2018multicarrier,uysal2019phase}.
To enhance the anti-interference capability of ISAC signal in the low SNR regime, ISAC signals combining OFDM and spread spectrum are proposed \cite{nusenu2018time,kim2012novel,cheng2015spread,sharma2020multicarrier}.
The characteristics of OFDM-based ISAC signals are summarized in Table \uppercase\expandafter{\romannumeral2}.

\subsection{ISAC Signal Combining OFDM and LFM}
LFM signal is a frequency continuous signal applied in high-accurate radar sensing. In addition, its frequency changes with time. LFM is also called chirp because its frequency changes like a bird call.
The echo signal contains delay and Doppler frequency shift.
The distance and velocity of the target are estimated through signal processing.

The LFM signal is given by~\cite{4101252}
\begin{equation}
	s\left( t \right) = \exp \left( {j2\pi \left( {{f_0}t + \frac{1}{2}k{t^2}} \right)} \right){\rm{rect}}\left( {\frac{{t - {T_{LFM}}}}{{{T_{LFM}}}}} \right),
\end{equation}
where ${T_{LFM}}$ is the duration of LFM signal, ${f_0}$ is the starting frequency and $k$ is the chirp rate of LFM.

\begin{figure*}[!htbp]
	\centering
	\includegraphics[width=0.85\textwidth]{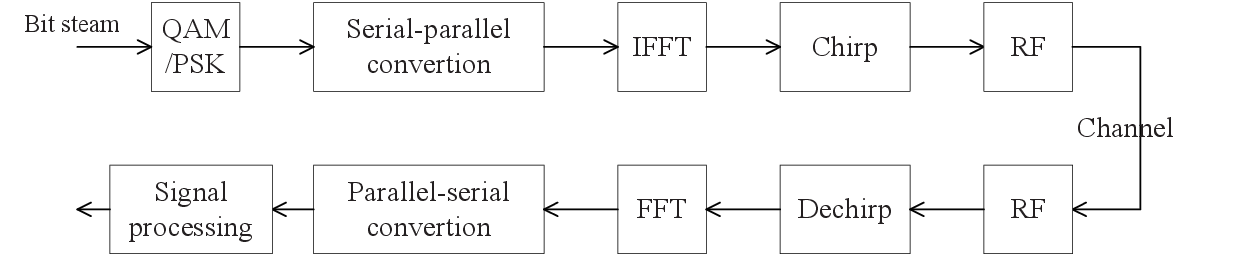}
	\caption{Signal processing procedure for OFDM-LFM signal \cite{4101252}.}
	\label{Signal processing procedures for OFDM-LFM signals}
\end{figure*}

The instantaneous frequency of the LFM signal increases with time, thereby increasing the time-bandwidth product of the signal and realizing high-resolution and long-distance radar sensing \cite{wang2012orthogonal}.
The ISAC signal combining OFDM and LFM, namely OFDM-LFM ISAC signal, effectively improves the Doppler sensitivity and reduce the estimation error of velocity \cite{li2018communication}.
The signal processing procedure for OFDM-LFM signals is shown in Fig. \ref{Signal processing procedures for OFDM-LFM signals}.
Compared with the signal processing procedure of OFDM, OFDM-LFM signals have chirp and dechirp processes.

According to the modulation methods, OFDM-LFM signals are classified into two types, namely carrier modulation and non-carrier modulation, as shown in Table \uppercase\expandafter{\romannumeral3}.
The OFDM-LFM signals based on carrier modulation modulate the symbols onto the LFM signals. Non-carrier modulation changes the OFDM-LFM signals, such as changing the sideband or subcarrier arrangement to transmit data \cite{takahashi2019novel,dash2015generalized}.

\subsubsection{Carrier modulation}
Darlington \emph{et al.} first patented LFM modulation in 1954 \cite{sidney1954pulse}.
Winkler \emph{et al.} improved LFM modulation in 1962 \cite{winkler1962chirp}.
In the early stage of LFM modulation, binary data 0 and 1 are mapped into positive and negative 
frequency modulation (FM) frequencies for data transmission.
This is approximately regarded as BPSK modulation, where 0 denotes that frequency increases over time, and 1 indicates that frequency decreases with time \cite{4101252}.

The data rate of the above methods is very low.
Therefore, high-order modulation schemes are applied to improve the data rate of LFM modulation, such as FSK \cite{wang2018joint,dwivedi2019target,striano2019performance}, PSK \cite{li2018communication,alabd2019partial} and QAM \cite{lv2018joint}.

The OFDM-LFM signal based on carrier modulation is expressed as \cite{wang2012orthogonal}
\begin{equation}
	\begin{aligned}
		&s\left( t \right) = \sum\limits_{m = 0}^{M - 1} {\sum\limits_{n = 0}^{N - 1}} ({d\left( {m N + n} \right)}\exp \left( {j2\pi {f_n}t} \right)\\
		&\quad \;\; \quad \;\;\cdot \exp \left( {j\pi {k}{t^2}} \right){\rm{rect}}( {\frac{{t - m {T_{sym}}}}{{{T_{sym}}}}} )).
		\end{aligned}
\end{equation}

To reduce the BER of communication and raise the resolution of distance estimation,
a modulation scheme using fractional fourier transform (FRFT) is proposed in \cite{chen2015study}.
Small phase BPSK modulation is applied to improve the concealment of the signal.
Compared with traditional BPSK modulation, the phase changes of small phase BPSK are $ \pm \varphi $, where $ 0 < \varphi  < {\raise0.7ex\hbox{$\pi $} \!\mathord{\left/
{\vphantom {\pi  2}}\right.\kern-\nulldelimiterspace}
\!\lower0.7ex\hbox{$2$}} $ \cite{zhang2017rf}.

\subsubsection{Non-carrier modulation}
\begin{table*}[htbp]
	\caption{ ISAC signals combining OFDM and phase coding }
	\renewcommand{\arraystretch}{1.7}
	\begin{center}
		\begin{tabular}{|m{0.22\textwidth}<{\centering}|m{0.15\textwidth}|m{0.55\textwidth}|}
			\hline
			\textbf{Category} & \makecell[c]{\textbf{References}} & \makecell[c]{\textbf{One sentence summary}}\\
			\hline
			Waveform modulated by phase coding sequence & \makecell[c]{\cite{tian2017waveform,hu2014radar,zhao2014chaos,qi2019phase}} & Phase coding sequence is modulated to subcarriers to reduce PAPR. \\
			\hline
			Phase modulated continuous waveform & \makecell[c]{\cite{zhang2017waveform,sahin2017novel,dokhanchi2018multicarrier,8904433}} & This method combines phase coding sequence and LFM. Phase coding sequence restrains PAPR and LFM provides good velocity estimation performance. \\
			\hline
		\end{tabular}
	\end{center}
\end{table*}

\begin{table*}[htbp]
	\caption{ Performance comparison of phase coding sequences}
	\renewcommand{\arraystretch}{1.7}
	\begin{center}
		\begin{tabular}{|m{0.18\textwidth}<{\centering}|m{0.28\textwidth}<{\centering}|m{0.285\textwidth}|m{0.15\textwidth}|}
			\hline
			\textbf{Phase code} & \textbf{Computational complexity of encoding} & \makecell[c]{\textbf{Sidelobe level of the }\\\textbf{autocorrelation function}} & \makecell[c]{\textbf{Envelope of signal}}  \\
			\hline
			Biphase code & Low & \makecell[c]{High} & \makecell[c]{Low} \\
			\hline
			Code coding & Medium & \makecell[c]{Medium} & \makecell[c]{Low} \\
			\hline
			Amplitude and phase modulation & High & \makecell[c]{Close to zero} & \makecell[c]{Medium} \\
			\hline
			Complementary code & High & \makecell[c]{Zero between symbols} & \makecell[c]{Low} \\
			\hline
		\end{tabular}
	\end{center}
\end{table*}

Takahashi \emph{et al.} design an ISAC signal to improve Doppler tolerance, where the communication signal is multiplexed at the sideband of an LFM pulse \cite{takahashi2019novel}.
The original LFM pulse is transformed into frequency domain by calculating its FFT, and part of the subcarrier signal in the sideband of the LFM pulse is replaced by complex symbols for communication.
Under the above modulation mode, there is no interference between communication and radar. The DIR is proportional to the number of communication subcarriers.

The large time-bandwidth product of the OFDM-LFM signal improves the sensing resolution.
To obtain the signal with a large time-bandwidth product and a suitable signal envelope, Dash \emph{et al.} propose a generalized OFDM-LFM signal design, which combines LFM and OFDM signals to achieve a high-resolution of distance and velocity estimation \cite{dash2015generalized}.
They interleave zeros in the input sequence to generate a new sequence. Then, other sequences are generated by shifting the new sequence. Therefore, the above sequences are orthogonal.
Assuming that the input sequence is ${x_p}[n]$, as shown in (\ref{eq4}), with $N$ discrete spectral components, the input sequence is interleaved by $K-1$ zeros between the two components to form. A new data sequence ${x_1}[n]$ is obtained with the above method.
Then ${x_1}[n]$, as shown in (\ref{eq5}), is cyclically shifted to generate ${x_2}[n]$\dots${x_K}[n]$, as shown in (\ref{eq6}). The new $K$ input sequences are calculated by $NK$ points IDFT to get the signals in time domain which occupy $NK$ subcarriers.
However, only the $N$ subcarriers are used to transmit data.
Since the input sequence is obtained by cyclic shift, the subcarriers used by data sequences, ${x_2}[n]$\dots${x_K}[n]$, are shifted with frequency spcaing of $\Delta f$.
The insertion of zero sequences reduces the peak power of the signal, reducing PAPR.
Two adjacent signals are distinguished according to the shift of the used subcarriers. However, the $NK$ subcarriers only carry $N$ modulation symbols.
\begin{equation}\label{eq4}
	{x_p}[n] = \{ x[0],x[1],x[2],..,x[N - 1]\},
\end{equation}
\begin{equation}\label{eq5}
	\begin{array}{l}
	{x_1}[n] = \{ x[0],{0_1},{0_2},...,{0_{K,}}x[1],{0_1},{0_2},...,{0_{K,}}...,\\
	\;\quad \quad \quad x[N - 1],{0_1},{0_2},...,{0_K}\}
	\end{array},
\end{equation}
\begin{equation}\label{eq6}
	\begin{array}{l}
	{x_K}[n] = \{ {0_1},{0_2},...,{0_{K,}}x[0],{0_1},{0_2},...,{0_{  K,}}x[1],...,\\
	\;\quad \quad\quad \quad {0_1},{0_2},...,{0_K},x[N - 1]\}
    \end{array}.
\end{equation}

\subsection{ISAC Signal Combining OFDM and Phase Coding}
\begin{figure*}[!htbp]
	\centering
	\includegraphics[width=1\textwidth]{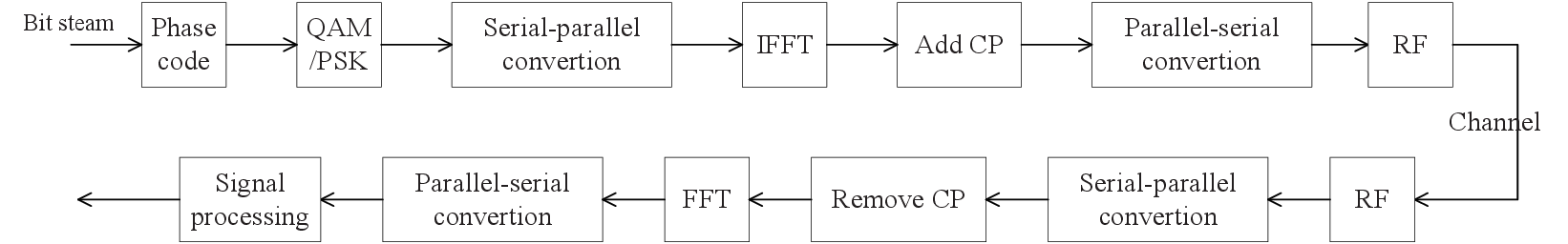}
	\caption{Signal processing procedure for phase-coded OFDM signal \cite{tian2017waveform}.}
	\label{fig5}
\end{figure*}
\begin{table*}[htbp]
	\caption{ ISAC signal combining OFDM and spread spectrum }
	\renewcommand{\arraystretch}{1.7}
	\begin{center}
		\begin{tabular}{|m{0.3\textwidth}<{\centering}|m{0.15\textwidth}<{\centering}|m{0.4\textwidth}|}
			\hline
			\textbf{Categories of spread spectrum sequences} & \textbf{References} & \makecell[c]{\textbf{One sentence summary}} \\
			\hline
			Gold sequence & \cite{kim2012novel,cheng2014mimo,ma2019integrated,cheng2015spread} & Gold sequence is the best coding sequence to reduce the peak of cross-ambiguity function. \\
			\hline
			Minimum-shift keying (MSK) sequence & \cite{liang2019design,tang2019analysis} & ISAC signal based on MSK sequence has a strong correlation.  \\
			\hline
			Multi-carrier direct sequence & \cite{sharma2020multicarrier} & ISAC signal based on multi-carrier direct sequence obtains ideal ambiguity function and low PAPR. \\
			\hline
		\end{tabular}
	\end{center}
\end{table*}
OFDM signal based on phase coding optimizes the peak
sidelobe ratio (PSLR), improving signal envelope performance
of the ambiguity function \cite{tian2017waveform}. The specific phase sequence
is modulated to the subcarriers of OFDM after constellation
mapping. Existing schemes are divided into two cases, namely,
direct phase coding sequence modulation and phase modulated
continuous waveform (PMCW). The combination of phase
coding and LFM, as shown in Table \uppercase\expandafter{\romannumeral4}, realizes a high time-bandwidth product and reduces PAPR.

\subsubsection{Direct phase coding sequence modulation}
A phase coding sequence with high autocorrelation reduces the sidelobe level of ambiguity function.
The selection of phase coding sequence affects the performance of radar sensing.
Commonly used phase coding sequences include biphase code, polyphase code, amplitude and phase modulation code (i.e., Huffman code), and complimentary code.
The comparison of these phase coding sequences is shown in Table \uppercase\expandafter{\romannumeral5}.

The OFDM signal combined with phase coding is
\begin{equation}\label{eq7}
	\begin{aligned}
		s\left( t \right) = \sum\limits_{n = 0}^{N - 1} {\sum\limits_{m = 0}^{M - 1} {{a_{nm}}  \exp \left( {j2\pi {f_n}t} \right){\rm{rect}}\left( {\frac{{t - m{t_b}}}{{{t_b}}}} \right)} },
	\end{aligned}
\end{equation}
where ${t_b}$ is the duration of the phase code, and ${a_{nm}}$ is the phase coding sequence \cite{tian2017waveform}.
The signal processing procedure for phase-coded OFDM signal is shown in Fig. \ref{fig5}.
Compared with OFDM, the bit stream is modulated with a phase code sequence prior to QAM/PSK.
A dephasing coding module is added at the communication RX.

ISAC signals based on phase coding have been studied extensively \cite{hu2014radar,tian2017waveform,tian2017radar,zhao2014chaos,qi2019phase}.
Hu \emph{et al.} apply random sequences to design ISAC signals that meet the requirements of radar and communication \cite{hu2014radar}.
The deserialized digital sequence is modulated in the corresponding subcarriers, and the information required for radar and communication are extracted at the RX, respectively.

A phase-coded OFDM signal design for a single scattering point target is proposed in \cite{tian2017waveform}.
The OFDM signal comprises $N$ subcarriers, and the bit stream is mapped onto an $M$-bits sequence.
They also propose a radar signal processing algorithm that achieves distance and velocity estimation with high-resolution and high-DIR \cite{tian2017radar}.

Traditional pseudo-random sequences, such as maximum length linear shift register sequences ($m$-sequence) and Gold sequences, have good autocorrelation.
However, their cross-correlation is not ideal. Moreover, the length of the $n$-th order $m$-sequences and Gold sequences is ${2^{n}} - 1$.
They have a limited number of sequence types and cannot be selected
randomly. Therefore, a large number of literatures have paid
attention to chaotic sequences, which are pseudorandom and
only related to the initial state and free parameters.
Zhao \emph{et al.} design a phase-coded OFDM signal using chaotic sequences, which has a large time-bandwidth product \cite{zhao2014chaos}.
They propose a method to extract phase code sequences from chaotic sequences with a good correlation.
Then, they use the method of subcarrier weighting to reduce the PAPR of the signal. Besides, the signal design has high flexibility to support different application scenarios \cite{zhao2014chaos}.
Recently, Qi \emph{et al.} propose the ISAC signal combining OFDM and phase coding with complete complementary codes, which achieves a high DIR, and accurate target detection \cite{qi2019phase}.

\subsubsection{Phase modulated continuous waveform (PMCW)}
PMCW provides high-resolution sensing. The ambiguity function of PMCW is sharp pushpin type, 
so that the distance-Doppler coupling is reduced \cite{zhang2017waveform,sahin2017novel}.
Moreover, PMCW is easily implemented.

Dokhanchi \emph{et al.} propose an automotive ISAC system based on a 
multicarrier-phase modulated continuous waveform (MC-PMCW). 
MC-PMCW reduces the modulation complexity of PMCW signal. 
Due to the PMCW, the proposed signal improves the anti-noise and 
anti-interference performance of OFDM signal \cite{dokhanchi2018multicarrier}.

In the scenario of MTC, 
the multiplexing strategy is applied to ISAC system to improve the identifiability of parameters.
Dokhanchi \emph{et al.} embed communication symbols into PMCW to reduce 
distance-Doppler ambiguity.
At the TX, the sampled sensing and communication sequences are modulated with 
differential phase shift keying (DPSK) in frequency domain and transformed to 
time domain through IFFT \cite{dokhanchi2019performance,dokhanchi2019mmwave}.

ISAC signal based on LFM or phase coding alone cannot satisfy the extremely high 
requirements of sensing and communication.
Hence, combining phase coding and LFM has attracted widespread attention \cite{uysal2019phase}.
Related designs appear in the ISAC-enabled automotive radars.
In \cite{uysal2019phase}, a group delay filter is
applied to process radar echo signals with inconsistent phase
coding. The signal-to-interference ratio (SIR) is efficiently
enhanced. The RX uses stretching processing to reduce the
sampling requirements of the received signal. This method
allows the RX to preserve phase coding through dechirp
processing because of the transmitted phase-encoded LFM
signal.

\subsection{ISAC Signal Combining OFDM and Spread Spectrum}
\begin{figure*}[!htbp]
	\centering
	\includegraphics[width=1\textwidth]{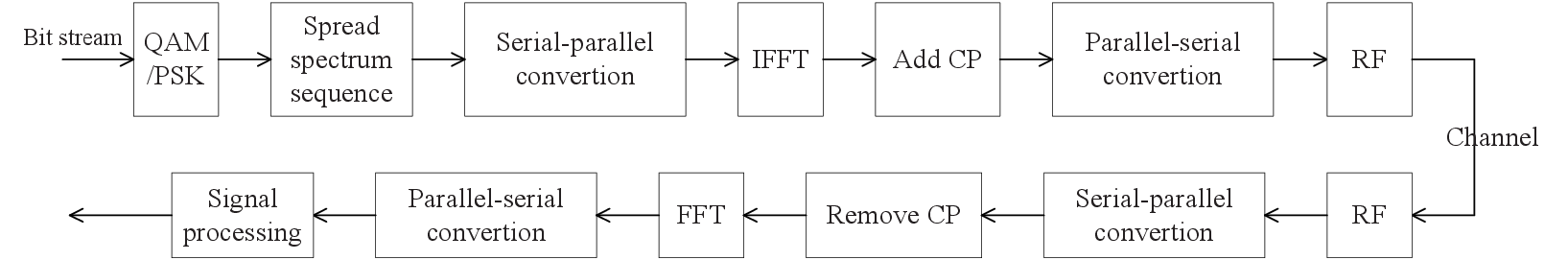}
	\caption{Signal processing procedure for SS-OFDM signal \cite{ma2019integrated}.}
	\label{Signal processing procedures for SS-OFDM signals}
\end{figure*}
Spread spectrum technology has the advantages of high confidentiality,
low power spectral density and anti-interference capability.
The combination of OFDM and spread spectrum, namely
spread spectrum-OFDM (SS-OFDM), reduces the PAPR of
OFDM signal. Among the spread spectrum technologies,
DSSS is widely applied in the design of ISAC signals. In
addition, the selection of spread spectrum sequences directly
impacts the performance of ISAC signals, as shown in Table \uppercase\expandafter{\romannumeral6}.

The implementation process of SS-OFDM is provided as follows. The communication data are first modulated at the
TX and then spread with a spread spectrum sequence. Then,
the CP is added after IFFT. At the RX, with the processes
of FFT, demodulation, and despreading, the original
communication data are recovered \cite{nusenu2018time,shen2016research}.
The SS-OFDM signal is given by \cite{ma2019integrated}
\begin{equation}
	\begin{aligned}
		s\left( t \right) =& \sum\limits_{m = 0}^{M - 1} {\sum\limits_{n = 0}^{N - 1} {\sum\limits_{p = 0}^{P - 1} {} } } (d\left( {n,m} \right){a_m}\left( p \right)\exp \left( {j2\pi {f_n}t} \right)\\
		& \cdot {\rm{rect}}( {\frac{{t - p{t_c} - m{T_{sym}}}}{{{T_{sym}}}}} ) ),
		\end{aligned}
\end{equation}
where $P$ is the length of the spread spectrum sequence. ${a_m}\left( p \right)$ is the spread spectrum code, with the value of $\pm 1$. ${t_c}$ is the duration of a spread spectrum code.
The signal processing procedure for SS-OFDM signals is shown in Fig. \ref{Signal processing procedures for SS-OFDM signals}.
Compared with OFDM, the modulation symbols are multiplied by the spread spectrum sequence.
A despread spectrum module is added at the communication RX.

The follow-up studies have shown that the spread factor affects the sidelobes of the ranging ambiguity function. Gold sequence is a commonly used spread spectrum sequence, which efficiently reduces the peak of the cross-ambiguity function \cite{cheng2014mimo,ma2019integrated}.
The OFDM signal combining spread spectrum and LFM is proposed to improve the resolution of the ISAC signal. The signal design ensures the orthogonality of the signal at the RX and has a large time-bandwidth product \cite{cheng2015spread}.

In addition to the Gold sequence, ISAC signals using minimum-shift keying (MSK) and DSSS are designed in \cite{liang2019design,tang2019analysis}.
The MSK signal has the advantage of strong correlation \cite{tang2019analysis}.
Besides, the optimization method of phase coding at the TX and the method to increase the PSLR at the RX are proposed, which effectively suppress the sidelobe level with sacrificing DIR.

The ISAC signal using multi-carrier direct sequence is proposed to obtain an ideal ambiguity function and low PAPR \cite{sharma2020multicarrier}.
The subset of $M$ subcarriers uses scrambling codes and orthogonal variable spread factor codes to reduce the correlation between subcarriers. Communication and radar functions are isolated due to the application of scrambling codes, reducing interference between radar and communication.

\subsection{ISAC Signals Using the Candidate Signals of 5G-A/6G}
FBMC, GFDM, DFT-s-OFDM and OTFS are proposed to solve the problems of Doppler sensitivity, high CP overhead and high PAPR of OFDM \cite{gerzaguet20175g}. Another candidate signal for the future mobile communication system, namely universal filtered multi-carrier (UFMC), has a higher OOB than OFDM, GFDM and FBMC, which is rarely used to design ISAC signals. In this section, we introduce the ISAC signals using the candidate signals of 5G-A/6G.

\subsubsection{FBMC}
Different from the complex input data of OFDM, the data stream of FBMC applies real data. FBMC applies prototype filters to ensure orthogonality between subcarriers, thereby avoiding the overhead of CP \cite{cao2016feasibility}.
The expression of the FBMC signal is
\begin{equation}
	\begin{aligned}
		s\left( t \right) =& \sum\limits_{n = 0}^{N - 1} {\sum\limits_{m = 0}^{M - 1} ({d\left( {mN + n} \right){g_f}\left( t \right)} } \\
		&\cdot \exp \left( {j2\pi {f_n}t} \right){\mathop{\rm rect}\nolimits} ( {\frac{{t - m{T_{sym}}}}{{{T_{sym}}}}} )),
		\end{aligned}
\end{equation}
where ${g_f}\left( t \right)$ is the prototype filter.

In terms of communication, FBMC has no CP and achieves higher spectrum efficiency than OFDM.
In terms of radar sensing, FBMC has a pushpin ambiguity function.
Due to the characteristics of large bandwidth, FBMC attains the distance resolution equivalent to OFDM.
In addition, FBMC does not have the sidelobes brought by CP, and its Doppler bandwidth is larger than OFDM, thereby improving the estimation performance of distance and velocity \cite{koslowski2014using}.
The orthogonality between subcarriers is destroyed by the filters.
Therefore, the QAM symbol carried by the FBMC signal is not easy to be demodulated.
Moreover, the bandwidth of the filter is narrow and there is a long tail in time domain,
which is also one of the disadvantages of FBMC signals.

\subsubsection{GFDM}
\begin{figure*}[!htbp]
	\centering
	\includegraphics[width=0.8\textwidth]{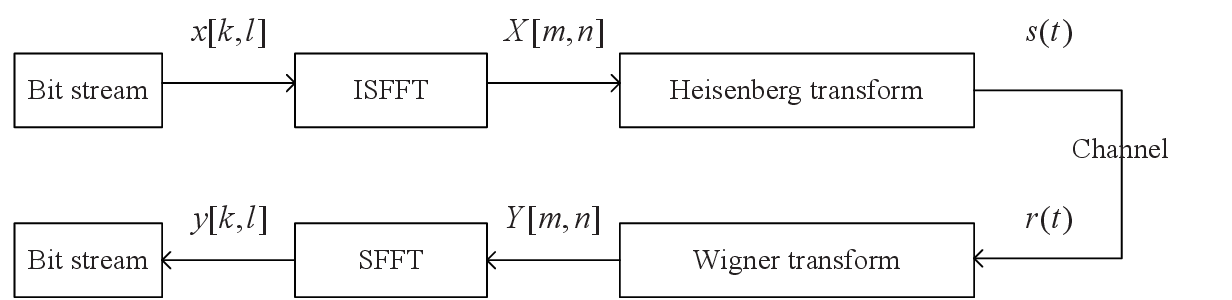}
	\caption{Signal processing procedure for OTFS signal \cite{raviteja2018interference}.}
	\label{Signal processing procedures for OTFS signals}
\end{figure*}
GFDM divides subcarriers into multiple groups and adds a filter to each group. The filter of GFDM is wider in frequency domain than that of FBMC, so that the tail of the filter in time domain is shorter.
The GFDM signal is 
\begin{equation}
	\begin{aligned}
		s\left( t \right) =& \sum\limits_{n = 0}^{N - 1} {\sum\limits_{m = 0}^{M - 1} ( {d\left( {mN + n} \right){g_{k,GFDM}}\left( t \right)} } \\
		&\cdot \exp \left( {j2\pi {f_n}t} \right){\mathop{\rm rect}\nolimits}( {\frac{{t - m{T_{sym}}}}{{{T_{sym}}}}} )),
		\end{aligned}
\end{equation}
where ${g_{k,GFDM}}\left( t \right)$ is the filter used for the $k$-th group of subcarriers, which has wide bandwidth to support multiple subcarriers \cite{michailow2012generalized}.
The filter of GFDM is simpler to implement compared with that of FBMC. The modulation method of the GFDM signal is to modulate block-like time-frequency resources.
Compared with other signals, it has a more flexible frame structure. 
However, the complex receiving algorithm is required to eliminate 
inter-symbol interference (ISI) and ICI.
The filters preserve the circular property of the signals in both time and frequency domains \cite{michailow2012generalized}. 
Therefore, the severe mutual-interference is avoided and the OOB is reduced. 
The performance of communication and sensing using OFDM and GFDM signals is analyzed 
in \cite{sanson2019non}, 
where GFDM achieves higher distance resolution and reduces 
mutual-interference compared to OFDM \cite{sanson2019non}.

\subsubsection{DFT-s-OFDM}
Compared with the OFDM-based ISAC signal, the DFT-s-OFDM based ISAC signal
has lower PAPR and simple implementation, while retaining the main advantages of OFDM,
which allows the receiver to use multi-user joint processing
frequency division multiple access (FDMA) and frequency domain equalization,
while reducing the PAPR of each user, so that it obtains large coverage \cite{ha2019post}.
In the DFT-s-OFDM signal, data symbols are expanded by DFT blocks,
and then undergo IDFT \cite{lee2020sub,huq2019terahertz}.
In order to avoid ISI caused by the
multi-path of the channel and allow frequency domain equalization at the receiver,
CP is set to the beginning of the symbol in advance. 
By changing the size of
the DFT expansion block, block-based single carrier
signals with different bandwidths are synthesized.
It is proved that DFT-s-OFDM signal contains a flexible internal protection cycle,
which does not affect the symbol duration \cite{berardinelli2017generalized}.

Meanwhile, the DFT-s-OFDM ISAC signal is combined with MIMO to achieve high
DIR \cite{shi2017papr}.
However, this scheme still generates high PAPR with
high-order QAM modulation. Considering the extremely
high data rate requirements of 5G-A and 6G, 
the high-order modulation is commonly applied.
Hence, how to maintain the low PAPR and
improve the spectral efficiency
of DFT-s-OFDM ISAC signal is the challenge of ISAC signal design.

\subsubsection{OTFS}
\begin{figure*}[!htbp]
	\centering
	\includegraphics[width=0.75\textwidth]{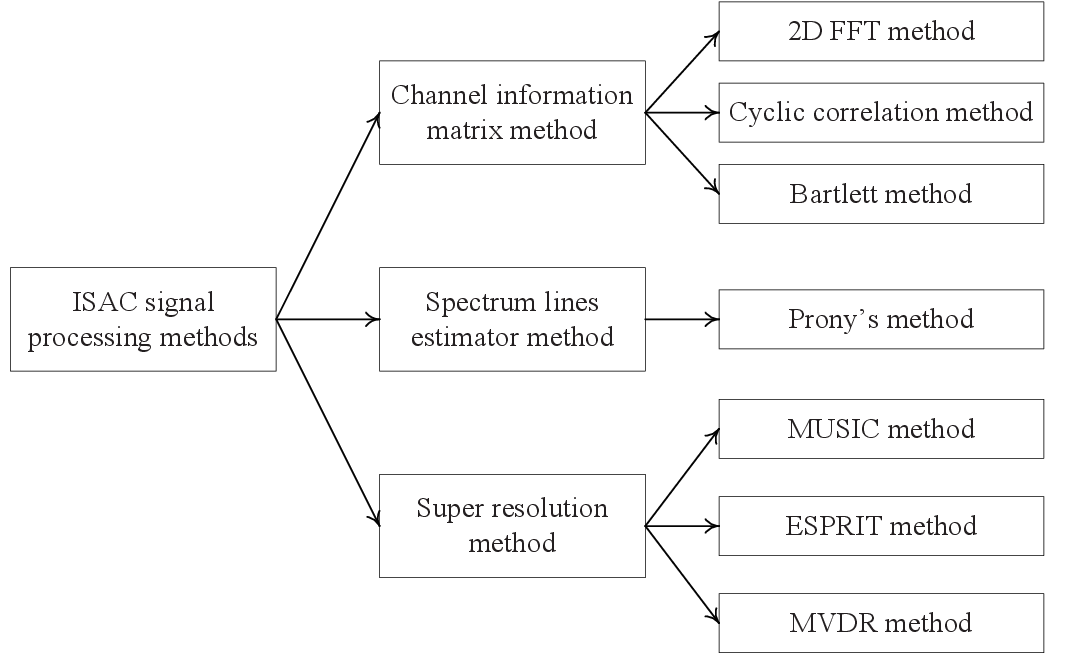}
	\caption{ISAC signal processing methods.}
	\label{Signal processing method}
\end{figure*}
OTFS is a two-dimensional (2D) modulation scheme designed in the delay-Doppler domains. 
The OTFS modulation
converts the dual dispersion channel into the delay-Doppler
domains using Heisenberg transform, which is approximately
equivalent to a non-fading channel \cite{hadani2017orthogonal,gaudio2019performance}.
OTFS modulation has high communication reliability and
efficiency in the environment with high mobility \cite{raviteja2018interference}, because the symbol in a frame experiences nearly the same constant
fading in the delay-Doppler domains.

OTFS is an extension of OFDM and is compatible with OFDM by adding a precoding module. The OTFS modulation process is shown in Fig. \ref{Signal processing procedures for OTFS signals}. 
In the first stage, the symbols in the delay-Doppler domains, namely ${x}[k,l]$, 
are converted into symbols in the time-frequency domains, namely $X[{m},{n}]$, 
through inverse symplectic finite Fourier transform (ISFFT). 
ISFFT is expressed as \cite{raviteja2018interference}
\begin{equation}
	X[{m},{n}] = \sum\limits_{k = 0}^{M - 1} {\sum\limits_{l = 0}^{N - 1} {{{x}[k,l]}\exp \left( {j2\pi \left( {\frac{{mk}}{M} - \frac{{nl}}{N}} \right)} \right)} } ,
\end{equation}
where $m,n$ and $k,l$ are the indices of the symbols in time-frequency domains 
and delay-Doppler domains, respectively.

Then, the obtained symbols are modulated by Heisenberg transform to form transmitted signal ${s}\left( {t} \right)$ \cite{raviteja2018interference}.
\begin{equation}
	\begin{aligned}
	&{s}\left( {t} \right) = \sum\limits_{{\rm{n}} = 0}^{N - 1} {\sum\limits_{m = 0}^{M - 1} ( {X\left[ {m,n} \right]} \exp \left( {j2\pi n\Delta f t} \right)} \\
	&\quad \;\;\quad \;\; \cdot {{\rm{rect}}}( \frac{{t - m{T_{sym}}}}{{{T_{sym}}}} ))
	\end{aligned},
\end{equation}

The signal in the delay-Doppler domains is obtained by calculating the IDFT and Wigner transform of the received signal at the RX.
Finally, the window function is applied to improve the channel sparsity in the delay-Doppler domains \cite{raviteja2019orthogonal}.

OTFS fully reveals the sparsity of wireless channels
using delay-Doppler impulse response in terms of communication.
Therefore, the number of pilots for channel estimation is
reduced, which improves the efficiency of channel estimation,
especially for the fading channel with large Doppler frequency shift.
Therefore, OTFS is crucial for communication in a highly
mobile environment \cite{raviteja2019orthogonal}.

In terms of radar sensing, OTFS is a fully digitally modulated signal.
Gaudio \emph{et al.} study the radar sensing performance of OTFS, OFDM, and FMCW.
OTFS and OFDM achieve the same sensing performance as FMCW, while transmitting at high DIR \cite{gaudio2019performance}.
Compared with OFDM, OTFS requires a shorter CP and achieves longer sensing distance and faster target tracking rates.
Moreover, OTFS is ICI-free, which enables large Doppler frequency shift estimation, and detects target with high velocity \cite{raviteja2019orthogonal}.

OTFS-based ISAC signal is first proposed in \cite{yuan2021integrated}.
TX transmits signals to RX according to the mobility state of RX, which is obtained by echo signal, to reduce the adverse effects of the channel.
The RX demodulates the communication data directly without channel estimation.
In other words, the entire OTFS frame is used for communication without setting the pilot used for channel estimation.

\section{ISAC Signal Processing}
The main radar signal processing methods for ISAC signals are divided into three categories, as shown in Fig. \ref{Signal processing method}, i.e., channel information matrix method, spectrum lines estimator method and super resolution method.
Channel information matrix method obtains sensing information through constructing a channel information matrix, including 2D FFT method, cyclic correlation method, and Bartlett method.
The representative method of spectrum lines estimator method is Prony's method.
Super resolution method estimates the distance and velocity of the target by decomposing the received signal into a noise subspace and a signal subspace, including  
multiple signal classification (MUSIC) method, 
estimating signal parameter via rotational invariance techniques (ESPRIT) method, 
and minimum variance distortionless response (MVDR) method.

In 2011, Sturm \emph{et al.} propose a radar signal processing algorithm applying 2D FFT to the received modulation symbol matrix, which simplifies radar signal processing \cite{sturm2011waveform}.
In order to improve the sensing accuracy and improve the anti-noise capability with low SNR,
the cyclic correlation method and Bartlett's method are proposed respectively \cite{wu2021integrating,oppenheim1978applications}. The cyclic correlation method is also
applicable to ISAC signals using OTFS.
Prony's method is proposed in 1795 \cite{prony1795essai}.
The sample space is obtained by oversampling the symbols of the OFDM signal.
Then, the sensing information such as distance and velocity, is obtained by solving the differential equation and the $n$-order polynomial of the pole.
Schmidt \emph{et al.} propose the MUSIC method in 1986 \cite{schmidt1986multiple}.
Firstly, the signals received at the antenna array are decomposed into signal subspace and noise subspace through singular value decomposition (SVD).
Then, the sensing information is obtained by eigenvalue estimation.
In 1986, Roy \emph{et al.} proposed the ESPRIT method, which applies the shift-invariant structure in the cisoids signal subspace. Besides, the sensing information is estimated by subspace decomposition and generalized eigenvalue calculation \cite{roy1986esprit}.
MVDR method is proposed by Capon in 1969, which has lower computational complexity than that of MUSIC and ESPRIT \cite{capon1969high}.
In the following subsections, these radar signal processing algorithms are reviewed.

\subsection{Performance Metrics of ISAC Signal Processing}
This section summarizes the performance metrics of ISAC signal processing, 
which are used to evaluate
the performance of ISAC signal processing in various scenarios.

\begin{itemize}
	\item[$\bullet$] \textbf{Computational complexity}: Computational complexity mainly refers to the 
	time consumption of ISAC signal processing, which is crucial for ISAC signal processing.
\end{itemize}

\begin{itemize}
	\item[$\bullet$] \textbf{Accuracy}: The accuracy reveals the difference between the sensing result
	and the real value.
\end{itemize}

\begin{itemize}
	\item[$\bullet$] \textbf{Root mean square error (RMSE)}: RMSE shows the error of
	ISAC signal processing algorithm.
	Since SNR affects the sensing results,
	RMSE also reveals the anti-noise performance of ISAC
	signal processing algorithm.
\end{itemize}

\begin{itemize}
	\item[$\bullet$] \textbf{Cramer-rao lower bound (CRLB)}: CRLB is obtained by calculating the likelihood function of the observation value and the Fisher information matrix.
	CRLB reveals the accuracy of radar sensing \cite{scharf1993geometry}.
\end{itemize}

\subsection{Channel Information Matrix Method}
\subsubsection{2D FFT method}
\label{31a}
2D FFT method directly processes the transmitted and received modulation symbols, rather than operating on the received signals.
This approach exploits the time-frequency structure of the OFDM signal and has low complexity \cite{sturm2011waveform}.
The transmitted OFDM signal is written as
\begin{equation}
	\begin{aligned}
		s(t) =& \sum\limits_{m = 0}^{M - 1} {\sum\limits_{n = 0}^{N - 1} ( {{{\bf{s}}_{{T_x}}}(mN + n)} }\\
		&\cdot{e^{j2\pi {f_n}t}}{\rm{rect}}( {\frac{{t - m{T_{sym}}}}{{{T_{sym}}}}} ) ),
	\end{aligned}
\end{equation}
where ${{\bf{s}}_{{T_x}}}$ is the communication modulation symbol sequence.
\begin{equation}
	\begin{aligned}
		&{{\bf{s}}_{R_x}}(mN{\rm{ + }}n){\rm{ = }}{\bf{H}}(m,n){{\bf{s}}_{{Tx}}}(mN+n){\kern 1pt}\\
		&\quad\quad\quad\quad\quad\quad\cdot {e^{ - j2\pi {f_n}\frac{{2{R_r}}}{c}}}{e^{j2\pi m{T_{sym}}{f_{d,r}}}},
	\end{aligned}
\end{equation}
where ${{\bf{s}}_{R_x}}$ is the communication symbols demodulated at RX,
${\bf{H}}$ represents the channel matrix,
${R_r}$ stands for the distance of target,
and $f_{d,r}$ represents Doppler frequency shift.

With the known phase shift of the $n$-th subcarrier eliminated, the channel information matrix
${{\bf{s}}_g}$ is obtained as

\begin{equation}
	{{\bf{s}}_g}\left( {m,n} \right) = \frac{{{{\bf{s}}_{R_x}(mN{\rm{ + }}n)}}}{{{{\bf{s}}_{T_x}(mN{\rm{ + }}n)}}} = {\bf{H}}(m,n){({\bar {\bf{k}}_r} \otimes {\bar {\bf{k}}_d})_{m,n}},
\end{equation}
with $\otimes$ referring to Kronecker product. The terms ${\bar {\bf{k}}_r}$ and ${\bar {\bf{k}}_d}$ are as follows \cite{sturm2011waveform}.
\begin{equation}
	{\bar {\bf{k}}_r} = {\left[ {{e^{ - j2\pi {f_1}\frac{{{\rm{2}}{R_r}}}{{\rm{c}}}}},{e^{ - j2\pi {f_2}\frac{{{\rm{2}}{R_r}}}{{\rm{c}}}}},...,{e^{ - j2\pi {f_n}\frac{{{\rm{2}}{R_r}}}{{\rm{c}}}}}} \right]^T},
\end{equation}
\begin{equation}
	{\bar {\bf{k}}_d} = \left[ {1,{e^{j2\pi {T_{sym}}{f_{d,r}}}},...,{e^{j2\pi \left( {M - 1} \right){T_{sym}}{f_{d,r}}}}} \right].
\end{equation}

The Doppler frequency shift ${f_{d,r}}$ and the distance ${R_r}$ are obtained by DFT transform for each row and IDFT transform for each column of the channel information matrix ${{\bf{s}}_g}$.

Denoting the peak index of DFT on $n$-th row of ${{\bf{s}}_g}$ as $in{d_{s,n}}$, the
velocity of target $v$ is derived using (\ref{eq17}) \cite{sturm2011waveform}.
\begin{equation}\label{eq17}
	\begin{aligned}
		&in{d_{s,n}} = \left\lfloor {{f_{d,r}}{T_{sym}}M} \right\rfloor,\\
		&{{f}_{d,r}} = \frac{{2v{f_c}}}{c},
	\end{aligned}
\end{equation}
where $\left\lfloor \right\rfloor$ is the floor function.

Similarly, denoting the peak index of IDFT on $m$-th column of ${{\bf{s}}_g}$ as $in{d_{s,m}}$, the distance ${R_r}$ is derived using (\ref{eq18}) \cite{sturm2011waveform}.
\begin{equation}\label{eq18}
	in{d_{s,m}} = \left\lfloor {\frac{{2{R_r}B}}{c}} \right\rfloor,
\end{equation}
where $B$ is the bandwidth of ISAC signal.

\subsubsection{Cyclic correlation method}
The signal processing
method based on correlation method adopts a virtual cyclic
prefix (VCP) according to the sampling points \cite{wu2021integrating}.
The maximum detection distance is related to the length of VCP.
Therefore, the maximum detection distance is no longer limited by CP.
The number of subcarriers and symbols in ISAC signal influence
the sensing performance of 2D FFT method. However,
these parameters are limited by existing communication standards
and cannot be dynamically set according to the sensing requirements.
Meanwhile, compared with 2D FFT method,
cyclic correlation (CC) method is more suitable for processing
OTFS based ISAC signals, where the communication symbols
modulated on subcarriers may be 0.
In this case, if a 2D FFT is applied,
the sampling value at the RX contains noise.
Direct division leads to noise amplification that affects the
accuracy of distance and velocity estimation \cite{wu2021integrating}.

The CC method samples the echo signal and transmitted signal in time domain,
divides the samples into multiple continuous sub-blocks,
then generates the channel information matrix by calculating the correlation
between the samples of the echo signal and the samples of the transmitted signal.
The estimation results of distance and velocity
are obtained by processing the channel information matrix.
The essential difference between CC method and 2D FFT method is the approach 
of grouping sample points.
At the RX, the sampling points are grouped in time domain 
in CC method instead of the grouping of 
samping points in time-frequency domains in 2D FFT method.
The channel information matrix in CC method is obtained by 
multiplying the values of grouped sampling points by the conjugate of the values of 
sampling points at the TX.
With the obtained channel information matrix, 
the distance and velocity of target are obtained
using maximum likelihood (ML) and discrete fourier transform (DFT) methods, respectively \cite{zeng2020joint}.

\subsubsection{Bartlett's method}

The sensing accuracy of 2D FFT method varies with the number of FFT points,
falling to achieve stable results.
According to the correlation of variance,
Bartlett divides the signal sequence into several groups,
and calculates the power spectrum function of each segment \cite{ali2019minimizing}.
This method not only obtains stable and accurates sensing results,
but also effectively mitigates the impact of noise on sensing results \cite{dwivedi2015optimal}.

Using the channel information matrix ${{\bf{s}}_g}$,
Bartlett's method obtains the column vector yielding distance and the row vector yielding velocity.
The obtained vector is divided into several groups,
and the power spectrum function of each group is calculated.
Then, the power spectrum function of each group is accumulated and averaged to obtain 
the average periodic graph.
The peak is searched according to the average periodogram.
Then, the distance and velocity of target are calculated \cite{oppenheim1978applications}.

\subsection{Spectrum Lines Estimator Method}

\begin{table*}[!htbp]
	\caption{Performance comparison of super resolution methods}
	\label{Performance comparison of super resolution methods}
	\renewcommand{\arraystretch}{1.7}
	\begin{center}
		\begin{tabular}{|m{0.33\textwidth}<{\centering}|m{0.23\textwidth}<{\centering}|m{0.16\textwidth}<{\centering}|m{0.16\textwidth}<{\centering}|}
			\hline
			\makecell[c]{\textbf{Super resolution methods}} & \makecell[c]{\textbf{Computational complexity}} & \makecell[c]{\textbf{Accuracy}} & \makecell[c]{\textbf{Resolution}}\\
			\hline
			MUSIC & High & High & Medium \\
			\hline
			ESPRIT & Medium & Medium & High \\
			\hline
			MVDR & Low & Low & Low \\
			\hline
		\end{tabular}
		\\
	\end{center}
\end{table*}

In 1795, Prony proposed a method using a set of linear combinations of exponential functions to fit uniformly sampled data.
The sensing information is obtained by solving difference equations.
In the presence of noise, this method transforms the parameter estimation into an optimization problem to improve the sensing resolution.
The steps of the Prony's method are provided as follows \cite{prony1795essai}.

\textbf{Step} 1: Data vector is obtained by adopting a sampling scheme. Assuming that the $k$-th sampling symbol at the RX is expressed as
\begin{equation}
	{s_k} = {\sum\limits_{n = 0}^{N - 1} {{a_n}\left( {{e^{{j2}\pi \left( {{f_n} + {\sigma _k}} \right)\Delta t}}} \right)} ^{k - 1}} + {\omega_k} ,
\end{equation}
where $\Delta t$ is the sampling interval, and $\omega_n $ is the AWGN. The purpose of this step is to estimate ${e^{{j2}\pi \left( {{f_n} + {\sigma _k}} \right)\Delta t}}$ from the sampled value.

\textbf{Step} 2: The characteristic equations of the linear difference equations are constructed as
\begin{equation}
	{s_k} + \sum\limits_{i = 1}^N {{b_i}} {s_{k - i}} = \omega \left( k \right) + \sum\limits_{i = 1}^N {{b_i}} \omega \left( {k - i} \right)\;k = 1,2,...,N.
\end{equation}

\textbf{Step} 3: Characteristic roots are derived accordings to the characteristic equations.

\textbf{Step} 4: Doppler frequency shift and delay of the signal are derived according to the characteristic roots \cite{prony1795essai}.

Prony's method uses a set of sampled values to obtain the sensing information.
However, Prony's method is affected by the sampling interval and noise.
Total least squares-singular value decomposition (TLS-SVD) algorithm is used to reduce the amount of calculation and overcome the impact of noise on ISAC signal processing \cite{prony1795essai}.
The characteristic roots of the signal are obtained according to the singular value, and Doppler frequency shift and delay are derived using the characteristic roots.

\subsection{Super Resolution Method}
The accuracy of 2D FFT and CC methods is limited by the sampling rate. 
Super resolution methods effectively overcome the limitations of the sampling rate, 
and provide high estimation accuracy and resolution compared with channel information matrix method and spectrum lines estimator method.
The MUSIC method comprehensively has the best accuracy among the super resolution methods with high computational complexity.
The accuracy and computational complexity of the ESPRIT method are both medium \cite{sanson2018comparison}.
Although the implementation of the MVDR method is simple, it has low resolution and accuracy. 
The performance comparison among the super resolution methods 
in terms of computational complexity, 
accuracy and resolution is shown in Table \ref{Performance comparison of super resolution methods}.

\subsubsection{MUSIC method}
MUSIC method adopts the matrix eigenspace decomposition, which provides an asymptotically unbiased estimation of
a) the number of incident wavefronts;
b) the direction of arrival (DoA) and distance of the target;
c) the strength and cross-correlation between the incident signals;
d) the strength of noise and interference.
The steps of the MUSIC method are provided as follows \cite{schmidt1986multiple}.

\textbf{Step} 1: The signal received at the antenna array composing
of $M$ antenna elements is the combination of echo signals and noise.
It is represented as a vector $\textbf{X}$.
\begin{equation}
	\textbf{X}{\rm{ = }}\textbf{AF} + \textbf{W},
\end{equation}
where $\textbf{A}$ is a matrix of phase difference of the echo signal
on the antenna elements due to the delay of signal propagation,
$\textbf{F}$ represents the sampling value of the echo signals on the reference antenna array,
and $\textbf{W}$ represents the noise received by the antenna array.

\textbf{Step} 2: The signal subspace and noise subspace are obtained by calculating the autocorrelation matrix ${\textbf{R}_{\textbf{XX}}}$ of $\textbf{X}$.

\textbf{Step} 3: The spatial spectrum ${P_{MU}}\left( \theta  \right)$ is obtained by constructing orthogonal relation equation according to the orthogonal relation between the signal subspace and the noise subspace.

\textbf{Step} 4: Sensing information, such as DoA, distance and velocity etc.,
is obtained by searching the peak of the spatial spectrum function.

\begin{figure}[!htbp]
	\includegraphics[width=0.5\textwidth]{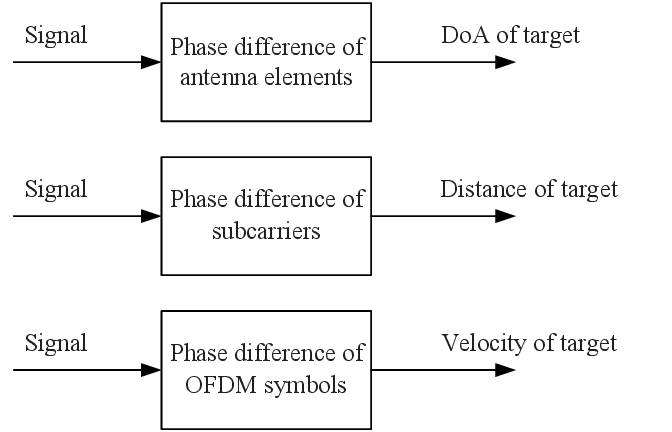}
	\caption{MUSIC method estimating DoA, distance and velocity of target.}
	\label{MUSIC}
\end{figure}

As shown in Fig. \ref{MUSIC}, when estimating DoA,
the echo signals received by different antenna elements are used to
construct spatial spectrum function.
As illustrated in Fig. \ref{MUSIC}, when estimating distance and velocity,
the channel information matrix is constructed as the method in Section \ref{31a}.
Then, the phase differences of subcarriers and OFDM symbols are applied to
estimate the distance and velocity of target, respectively.
Specifically, the spatial spectrum functions are
constructed using the row vector
and column vector of the channel information matrix by undergoing the above steps.
Then, the distance and velocity of target are 
estimated by searching the peaks of these spatial spectrum functions.
The MUSIC method provide parameter estimation close to CRLB.
Simulation results in \cite{schmidt1986multiple}
demonstrate that the MUSIC method estimates the DoAs of multiple targets with
higher accuracy than
ML \cite{5494616} and maximum entropy (ME) \cite{5161769} methods.

\subsubsection{ESPRIT method}
In order to solve the challenge of blind
synchronization of OFDM signal in mobile communication systems,
Ufuk \emph{et al.} propose ESPRIT \cite{tureli2000ofdm}.
This method exploits the
structure information of OFDM signals to estimate sensing
information.
Compared with MUSIC method, ESPRIT method
provides the same resolution with lower complexity.
Simulation results show that ESPRIT method is superior to MUSIC method 
in terms of complexity \cite{tureli2000ofdm}.

Similar to MUSIC method,
ESPRIT method exploits the signal and noise
to generate asymptotically accurate estimation of distance and velocity.
ESPRIT method has the following advantages.
a) ESPRIT method does not require the knowledge of antenna patterns.
b) The complexity of ESPRIT method is much lower than MUSIC method
because it does not need the search procedure inherent
in MUSIC algorithm, namely the Step 4 in MUSIC algorithm.
c) ESPRIT method simultaneously estimates the number of sources and their DoAs \cite{tureli2000ofdm}.

ESPRIT exploits the rotational invariance signal subspace.
The steps of ESPRIT method are provided as follows \cite{tureli2000ofdm}.

\textbf{Step} 1: The $K$ samples of the received signal are given by
\begin{equation}
	\textbf{X}\left[ k \right] = \sum\limits_{n = 1}^N {{a_n}{e^{j2\pi {f_n}t}}}  + \omega \left[ k \right], k = 0,1,...,K-1,
\end{equation}
where ${a_n}$ is the symbol modulated on the $n$-th subcarrier, and $\omega \left[ k \right]$ is the AWGN.

\textbf{Step} 2: In order to exploit the deterministic nature of the cisoids,
$\textbf{X} = \left[ {\textbf{X}\left[ 0 \right],\textbf{X}\left[ 1 \right],...,\textbf{X}\left[ K-2 \right]} \right]$
and $\textbf{Y} = \left[ {\textbf{X}\left[ 1 \right],\textbf{X}\left[ 2 \right],...,\textbf{X}\left[ K-1 \right]} \right]$ are constructed.
The autocorrelation matrix ${\textbf{R}_{\textbf{XX}}}$ and cross-correlation matrix ${\textbf{R}_{\textbf{XY}}}$ are derived.

\textbf{Step} 3: Calculating the eigendecomposition of autocorrelation matrix ${\textbf{R}_{\textbf{XX}}}$, the minimum eigenvalue is the variance of noise ${\sigma ^2}$.

\textbf{Step} 4: The generalized eigenvalue matrix ${\textbf{C}_{\textbf{XX}}}$ and ${\textbf{C}_{\textbf{XY}}}$ are calculated using the variance of noise as follows.
\begin{equation}
	{\textbf{C}_{\textbf{XX}}}{\rm{ = }}{\textbf{R}_{\textbf{XX}}} - {\sigma ^2}\textbf{I},
\end{equation}
\begin{equation}
	{\textbf{C}_{\textbf{XY}}}{\rm{ = }}{\textbf{R}_{\textbf{XY}}} - {\sigma ^2}\textbf{Z},
\end{equation}
where $\textbf{I}$ is the identity matrix, $\textbf{Z}$ is given by
\begin{equation}
	\textbf{Z} = \left[ {\begin{array}{*{20}{c}}
			0&.&.&.&0&1\\
			0&.&.&.&1&0\\
			.&{}&{}&.&.&.\\
			.&{}&.&{}&.&.\\
			.&.&{}&{}&.&.\\
			1&.&.&.&0&0
	\end{array}} \right].
\end{equation}

When estimating DoA,
the received signals of different antenna array elements
are used to construct the covariance matrices and the generalized eigenvalue matrices,
which are applied to calculate the subspace rotation operator.
The eigenvalue of subspace rotation operator yields DoA of target \cite{roy1986esprit}.
When estimating distance and velocity of target,
the channel information matrix is constructed as the method in Section \ref{31a}.
The covariance matrices and the generalized eigenvalue matrices are
obtained using the row vector
and column vector of the channel information matrix by undergoing the above steps.
Then, the subspace rotation operators for distance and velocity estimation are constructed,
whose eigenvalues yield the distance and velocity of target.

\subsubsection{MVDR method}

Capon \emph{et al.} propose MVDR method by in 1969.
Compared with MUSIC and ESPRIT methods, MVDR method has the 
advantages of easy implementation and minimizing the power of noise and interference,
thereby mitigating their impact on sensing performance.
MUSIC and ESPRIT  methods have better sensing performance than MVDR method with the same number of sampling points \cite{capon1969high}.

Similar to MUSIC and ESPRIT methods, MVDR method firstly constructs channel information matrix.
The column and row vector of the channel information matrix
are weighted summed to obtain the frequency response vectors 
projected into the delay and Doppler domains, respectively.
By using frequency response vectors, the autocorrelation matrices are constructed.
The spatial spectrum functions are then constructed according to the autocorrelation matrices 
and the peak indices of the spatial spectrum functions are obtained,
yielding the estimation of distance and velocity of target \cite{capon1969high}.

\subsection{Signal Processing for OFDM based ISAC Signals}
According to the above algorithms, 2D FFT estimates the
distance and velocity of target with low complexity. Therefore,
2D FFT method is widely applied to process most OFDMbased
ISAC signals. This section briefly reviews the signal
processing methods for the OFDM-based ISAC signals.

\subsubsection{ISAC signal combining OFDM and LFM}
The modulation symbol is recovered from the received signal by removing CP and undergoing FFT. The echo signal is expressed as
\begin{equation}
	\begin{aligned}
		r\left( t \right) =& \sum\limits_{n = 0}^{N - 1} {\sum\limits_{m = 0}^{M - 1} ( {{\bf{H}}\left( {n,m} \right)d\left( {mN + n} \right)} } {{\rm{e}}^{j2\pi {f_d}t}}\\
		& \cdot {{\rm{e}}^{j2\pi {f_n}\left( {t - \tau} \right)}}{{\rm{e}}^{j\pi {k_{r,m}}{{\left( {t - m{T_{sym}} - \tau} \right)}^2}}} ),
	\end{aligned}
\end{equation}
where ${\bf{H}}\left( {n,m} \right)$ is the channel response of the $n$-th subcarrier
of the $m$-th symbol, ${f_d}$ is the Doppler frequency shift, 
$\tau = \frac{{2{R_r}}}{c}$ is the delay \cite{wang2012orthogonal}.

At the RX, the same steps are performed in reverse order to the TX.
Because of the existence of CP, the demodulator can capture the signal of a complete symbol period and obtain the starting point of the symbol period, which guarantees that the result of 2D FFT is correct.
When the signal is dechirped, the received modulation symbols are recovered by calculating the FFT of the received baseband signal.

\subsubsection{ISAC signal combining OFDM and phase coding}
Combining OFDM and phase coding, as given in (\ref{eq7}), the received signal with down-conversion is obtained as \cite{tian2017waveform}
\begin{equation}
	\begin{aligned}
		&r\left( t \right) = \sum\limits_{n = 0}^{N - 1} {\sum\limits_{m = 0}^{M - 1} ( {{\bf{H}}\left( {n,m} \right){a_{n,m}}} }  \\
		&\quad \quad \quad\cdot {\rm{rect}}\left( {t - m{t_b} - \tau } \right){e^{j2\pi {f_n}\left( {t - \tau } \right)}} ).
	\end{aligned}
\end{equation}

There are multiple methods of ISAC signal processing.
Take 2D FFT method as an example.
The channel information matrix is obtained.
Then, the distance and velocity of target are obtained using 2D FFT method \cite{tian2017waveform}.

\subsubsection{ISAC signal based on spread spectrum}
Take 2D FFT
method as an example. Before the OFDM modulation of
the SS-OFDM signal, the communication
symbols are spread by the spread spectrum sequence.
Similarly, the received matrix is composed of the transmitted
symbols. The received matrix is divided
by the transmitted matrix, generating the channel information matrix. 
The distance and velocity of the
target are obtained using 2D FFT method \cite{chen2021code}.

\subsection{Signal Processing for OTFS based ISAC signals}
\begin{table*}[htbp]
	\caption{ISAC signal optimization}
	\label{tab8}
	\renewcommand{\arraystretch}{1.7}
	\begin{center}
		\begin{tabular}{|m{0.25\textwidth}<{\centering}|m{0.2\textwidth}<{\centering}|m{0.47\textwidth}|}
			\hline
			\textbf{Category} & \textbf{Optimization method} & \makecell[c]{\textbf{One-sentence summery}}\\
			\hline
			
			\multirow{3}[5]{\linewidth}{\makecell[c]{PAPR optimization}}
			& Coding method & The coding method is used to design the information sequence to avoid the occurrence of large signals. \\
			\cline{2-3}
			&Probability method & PTS and SLM are used to restrain PAPR.\\
			\cline{2-3}
			& Tone reservation & Part of LFM signals are used for information transmission, while other LFM signals are used for PAPR suppression.\\
			\hline
			
			\multirow{3}[6]{\linewidth}{\makecell[c]{Interference management}}
			& Mutual-interference cancellation & One user's ISAC signal brings interference to another user's reception of the echo of ISAC signal.\\
			\cline{2-3}
			& Self-interference cancellation &The signal sent by TX interferes with the echo signal received by RX.\\
			\cline{2-3}
			& Interference avoidance & Duplex or multiple access are used to avoid interference between users.\\
			\hline
			
			\multirow{5}[6]{\linewidth}{\makecell[c]{Adaptive signal optimization}}
			& Total power constraint &When the maximum power of the signal is determined, the ISAC signal is designed with the best performance.\\
			\cline{2-3}
			& Mutual information (MI) constraints & Design adaptive ISAC signals to achieve maximum MI.\\
			\cline{2-3}
			& Indoor environment & Adaptive ISAC signals are used to achieve optimal scattering characteristics.\\
			\cline{2-3}
			& Pilot structure design & The sensing performance and anti-noise ability of signal are improved by inserting pilot. \\
			\hline
		\end{tabular}
	\end{center}
\end{table*}
The RX of OTFS signal performs the inverse process of the TX, converting the received signal to the delay-Doppler domains for demodulation \cite{hadani2017orthogonal}. Specifically, the received signal
in the time domain is first processed by Wigner transform, and
then the symbols in the delay-Doppler domains are obtained by
adding a receiving window function and calculating SFFT as
shown in Fig. \ref{Signal processing procedures for OTFS signals}.

Gaudio \emph{et al.} propose OTFS-based ISAC signal and the
corresponding signal processing algorithm \cite{gaudio2020effectiveness}. 
The ML algorithm is applied in radar signal processing and 
the soft-output detector utilizing channel sparsity in 
Doppler-delay domains is applied for communication \cite{gaudio2020effectiveness}. 
Then, with the combination of OTFS and MIMO, they 
further applied ML algorithm in estimating the DoA, distance and velocity of target.
Two steps of coarse and refined parameter estimation are designed to reduce the complexity of ISAC signal processing \cite{gaudio2020joint}.
Raviteja \emph{et al.} propose a matched filter for the distance
and velocity estimation of the target. Compared with the CP
overhead in OFDM, OTFS requires less overhead. Meanwhile,
since OTFS signal has no ICI, it achieves more accurate
Doppler frequency shift estimation, achieving more accurate
velocity estimation compared with OFDM signal \cite{hadani2017orthogonal}.

\section{ISAC Signal Optimization}
ISAC signal optimization mainly includes three categories,
i.e., PAPR optimization, interference management, and adaptive
signal optimization, as shown in Table \ref{tab8}. ISAC signal with high PAPR is easy to enter
the non-linear region of the power amplifier, which leads to
signal distortion. Therefore, an effective PAPR reduction scheme
is necessary to reduce the PAPR of ISAC signal.
The RX will encounter the self-interference from the TX and the mutual-interference
from other TX, which will degrade the sensing performance. 
Hence, interference management is essential in
ISAC signal optimization. As the application scenarios of the ISAC signal are various, adaptive signal optimization is required,
including signal parameter optimization and signal structure
optimization. Through signal optimization, ISAC signal adapts
to various application scenarios, and improves the performance
of sensing and communication.

\subsection{PAPR Optimization}
High PAPR may cause OOB or in-band distortion of the transmitted OFDM signals \cite{han2004papr}. Thus, the OFDM signals require a large linear dynamic range of power amplifiers to avoid distortion. This subsection reviews coding method,
tone preserving method,
and probability method to reduce PAPR.

\subsubsection{Coding method}
When the OFDM signal has a large amplitude, it may produce a high PAPR. However, this only occurs when the modulation symbols of the OFDM signal form a tightly structured symbol sequence.
The coding method generates a signal with low PAPR by coding sequences \cite{tigrek2009relation}.
The bit stream is processed through the coding sequences to generate a plurality of candidate sequences.
The candidate sequences are modulated into symbol sequences respectively.
The PAPR of the OFDM signal for each symbol sequence is first calculated, and then the candidate sequences that produce high peak power are discarded.
Since some candidate sequences are abandoned, the bit stream needs to be transmitted with long codes.
For example, 3-bit information needs to be encoded with 4-bit coding sequences.
The coding methods have a large amount of calculation, because it needs to exhaust all the candidate sequences.
However, the coding methods are applicable to the case where the coding sequence is short and the number of subcarriers is small.
The number of candidate sequences that need to be exhausted will grow rapidly if there are many subcarriers \cite{li2019waveform}.

Take the Golay code as an example.
Golay code is a typical PAPR reduction coding method with strong error correction capability and high sensing performance \cite{tigrek2007golay}.
Using the Golay sequence, PAPR is reduced to 3 dB \cite{tigrek2009relation}.

The OFDM signal is a superposition of modulation symbols
over all the subcarriers. A self-disarrange block coding method
is proposed to reduce PAPR \cite{li2019waveform}. Firstly, Golay sequences are
applied to encode the information bits to obtain an encoded
packet. Then, the proposed algorithm generates all arrangements
of encoded packets on the OFDM symbol. Finally, the
modulation symbol sequence with the lowest PAPR is selected.
In addition, the P4 code is used to reduce PAPR, and the
phase code sequence of the signal is designed by cyclic
shift based on the P4 sequence \cite{tian2018hrrp}.

\subsubsection{Probability method}
The probability method is an extension
of the coding method. The idea of the probability method is to
use a sequence set to represent a bit stream of communication
information. Then, the sequence with the lowest PAPR is selected
from the sequence set and regarded as the transmission
sequence. Commonly used probability methods mainly include
partial transmission sequence (PTS) and selected mapping
(SLM) \cite{zhao2018judgement}.

The PTS divides the modulation symbols into several subblocks.
Each sub-block is modulated onto fixed subcarriers and
adjusted with different phase factors to minimize the PAPR of
each sub-block. The ability to reduce PAPR improves as the
number of sub-blocks increases \cite{zhao2018judgement}.

The SLM converts the original modulation symbols into low PAPR symbols.
This method generates modulation symbols according to the original modulation symbols created by the TX.
Then, the modulation sequence with a low PAPR is selected to transmit, thereby reducing PAPR without distorting the transmitted signal \cite{le2009selected}.

Both PTS and SLM need to record the side information (SI) indicating the modulation symbols.
The RX demodulates the modulation symbols according to the SI index to obtain the communication symbol.
The performance of communication and sensing at RX depends on the accuracy of SI \cite{zhao2018judgement,le2009selected}.

An SLM method without sending SI indices is proposed in
\cite{le2009selected}, where the SI indices are embedded into the modulation
sequence to reduce the overhead in the transmitted sequence.
In order to reduce PAPR efficiently, a method combining
SLM and PTS is proposed in \cite{zhao2018judgement}. The modulation symbol
sequences first undergo SLM. When the PAPR of the output
sequence of the previous step is smaller than a threshold, the
sequence is chosen to be modulated on the ISAC signal at TX.
Otherwise, the output sequence is further processed using the
PTS method. This method reduces the PAPR with low complexity
without degrading the performance of communication
and sensing.

\subsubsection{Tone reservation}
Tone reservation method reserves
some subcarriers to modulate LFM signals and generate
peak canceling signals. Other subcarriers are used to transmit
communication signals. This method has low complexity and
BER. However, some subcarriers are reserved to reduce PAPR,
resulting in the waste of spectrum resources.

In \cite{lv2018novel}, a PAPR reduction method is proposed using the tone reservation method.
An orthogonal chirp division multiplexing (OCDM) signal is composed of $N$ 
orthogonal LFM signals and OFDM signals.
Compared with classical tone reservation
method, this scheme requires fewer peak cancellation signals to
reduce PAPR. The performance of PAPR reduction improves
as the number of peak cancellation signals increases.
The tone preserving method is combined with the SLM
method by first performing the SLM method, and then performing
the tone preserving method, which reduces the PAPR
more effectively than the tone reservation method \cite{rema2017mimo}.

\subsection{Interference Management}
\begin{figure*}[!htbp]
	\centering
	\includegraphics[width=1\textwidth]{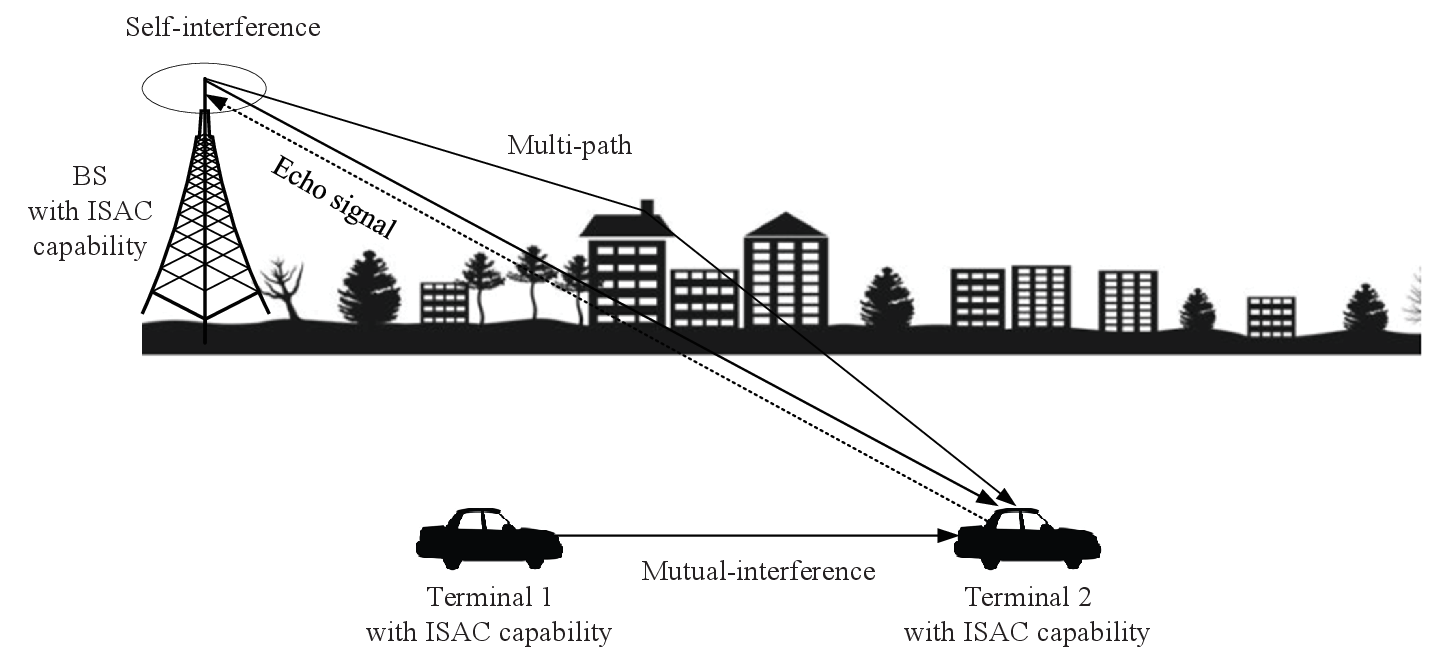}
	\caption{Multi-path and multi-user scenario.}
	\label{Multi-path and multi-user scenario}
\end{figure*}
Interference is one of the non-negligible factors that restrict
the performance of ISAC system.
ISAC interference mainly
consists of mutual-interference and self-interference. For the ISAC
system, since the echo signal usually returns to the RX
before the transmission is completed, the transmission will
interfere with the echo signal, which causes self-interference.
When the ISAC system is applied in the multi-path multi-user
scenario, the ISAC signal transmitted by one user may cause
interference to other users, resulting in mutual-interference.
If there is strong mutual-interference and self-interference, as
shown in Fig. \ref{Multi-path and multi-user scenario},
the sensing performance of ISAC system
will be severely degraded, failing to correctly identify targets
within its detection distance.

\subsubsection{Mutual-interference cancellation}
Typical mutual-interference cancellation algorithms consist of serial interference cancellation \cite{sit2011interference} and selective interference cancellation \cite{sit2012one,sit2011extension,sit2017mutual}.
\begin{itemize}
    \item[$\bullet$] Serial interference cancellation
\end{itemize}

The core method of the serial interference cancellation is to
demodulate the received ISAC signal. Then the reconstructed
interference signal from the received signal is subtracted \cite{sit2011interference}.
Schmidl and Cox (S${\rm{\& }}$C) algorithm is first used to detect interference signals through amplitude attenuation and phase rotation.
Then, starting from the identified interference signal
with the maximum power, the interference signal is decoded
successively. Finally, in order to reconstruct the signal, it is
necessary to remodulate the correct demodulation interference
signal, and add the amplitude attenuation, phase rotation,
frequency offset and delay. Then the reconstructed interference
signal is subtracted from the original ISAC echo signal
to eliminate the interference.

One of the challenges of serial interference cancellation
is that the amplitude attenuation and phase rotation of the
interference signals with the same delay cannot be detected by
correlation. It is shown in \cite{sit2011interference} that when the power difference between the identified interference signal and other received signals is smaller than 15 dB, the frequency offset may be
misestimated by S${\rm{\& }}$C algorithm. Moreover, the residual frequency offset in the estimation process will have a cumulative effect on the subsequent signal subtraction \cite{sit2012one,sit2011extension,sit2017mutual}.

\begin{itemize}
    \item[$\bullet$] Selective interference cancellation
\end{itemize}

Difference from serial interference cancellation, only the strongest identified interference will be reconstructed for selective interference cancellation \cite{sit2012one}. Selective interference cancellation only needs one cycle processing, and it does not leave ambiguous dots, which will not be mistaken as a target.

However, the performance of the two interference cancellation algorithms largely depends on the correct frequency offset estimation. Minor errors in frequency offset estimation may lead to incorrect signal reconstruction.

\subsubsection{Self-interference cancellation}

In addition to mutual-interference, self-interference is also the main factor impacting the performance of the ISAC system. The self-interference may be
much stronger than the echo signal, since the TX is close to the RX \cite{zeng2018joint}. Although the radar RX knows the transmitted
signal, the channel is random, making self-interference cancellation
challenging.

The adaptive interference cancellation method proposed in \cite{zhang2019adaptive} generates the cancellation signal with the opposite phase and equal amplitude to the self-interference signal, and then adds it to the received signal to cancel the self-interference.
In \cite{zeng2018joint}, instead of using a direct interference cancellation scheme,
the self-interference channel between transmitting and receiving antennas is
estimated to eliminate the self-interference and realize full duplex operation,
which is suitable for moving target detection.
In \cite{mazher2018automotive},
the self-interference of an omnidirectional antenna is considered,
and the least squares matching pursuit (LSMP) algorithm was used to
search all targets iteratively. \cite{barneto2019high} and \cite{barneto2019full}
indicate that promoting sufficient TX-RX isolation is a critical issue to be considered.
In order to achieve adequate TX-RX isolation, appropriately customized
RF and digital cancellation solutions are designed in \cite{barneto2019full}
to cancel the self-interference without
suppressing the echo signals of actual targets.

\subsubsection{Interference avoidance}
Another solution to interference management is interference avoidance, which adopts the schemes such as duplex or multiple access to avoid the interference, such that there is no need to apply complex interference cancellation schemes.
\begin{itemize}
    \item[$\bullet$] Mutual-interference avoidance
\end{itemize}

For multi-path and multi-user scenarios shown in Fig. \ref{Multi-path and multi-user scenario}, 
a multiple access scheme is applied to distinguish different ISAC users, 
so as to avoid mutual-interference. Common multiple access methods include time division multiple access (TDMA), code division multiple access (CDMA), 
orthogonal frequency division multiple access (OFDMA), 
and carrier sense multiple access (CSMA), 
which are applied to avoid mutual-interference in the ISAC system.

The principle of TDMA is that different ISAC users transceive signals within the specified time slots, avoiding the interference. In \cite{hellsten2020multiple}, the TDMA is applied for channel allocation among multiple radars to achieve the required channel isolation level.

With CDMA, ISAC users are assigned with orthogonal code
sequences for channel separation to avoid mutual-interference.
One method for mutual-interference avoidance is to combine
CDMA with DSSS techniques to design ISAC signal. In \cite{sharma2020multicarrier},
a new ISAC signal based on multicarrier-direct sequence-CDMA (MC-DS-CDMA) is proposed. 
The radar subsystem uses spread spectrum codes with ideal autocorrelation 
and cross-correlation characteristics to estimate distance, and uses scrambling codes 
to distinguish radar and communication signals. In \cite{lubke2020combining}, an ISAC signal combining radar sensing
with DSSS communication is proposed, which reduces the interference between multiple users. 
Another method is to adopt a signal processing scheme to suppress interference. 
In \cite{chen2021code}, 
a code division-OFDM (CD-OFDM) ISAC system is proposed. 
The corresponding serial interference cancellation-based signal processing method is 
proposed to suppress the mutual-interference, which achieves code division gain to 
suppress the interference between radar echo signal and communication signal.

With OFDMA, each user is assigned a subset of subcarriers,
and multiple users transmit simultaneously. In \cite{hadizadehmoghaddam2017simultaneous}, the concept of OFDMA radar is proposed, and a subset of subcarriers are allocated to each radar task so that the radar performs multiple tasks at the same time. OFDMA radar has the advantages of high sensing accuracy, strong anti-clutter and anti-interference ability, and high spectrum efficiency. In \cite{hadizadehmoghaddam2018interference}, based on OFDMA, the concept of interference-aware power allocation is proposed to improve sensing ability and
clutter suppression. The subcarriers overly affected by clutter will be dropped.

CSMA is a distributed multiple access method to avoid transmission conflict. In \cite{kurosawa2019proposal}, an FMCW radar based on CSMA is proposed, which simultaneously avoids both wideband and narrowband interferences causing miss detection and false alarm. In \cite{ishikawa2019packet}, a novel concept of packet-based FMCW radar using CSMA is proposed to reduce the probability of
narrowband interference causing false alarm of ghost target \cite{goppelt2010automotive}.

\begin{itemize}
    \item[$\bullet$] Self-interference avoidance
\end{itemize}

Self-interference is the main challenge of a full-duplex ISAC system, which will degrade the sensing performance. The performance of self-interference cancellation under 
the full duplex assumption is limited and has high computational complexity in the digital domain.
Therefore, self-interference avoidance is essential to improve the sensing performance with imperfect self-interference cancellation. In \cite{tang2020self}, the Golay sequence in the preamble is applied to reduce self-interference in short-distance sensing. In addition, for long-distance sensing,
a pilot signal is designed, 
where the OFDM symbols without self-interference at the end of OFDM data frame is uesd to 
achieve accurate distance estimation. This method does not need self-interference cancellation 
in the digital domain, which
reduces the computational complexity. In \cite{lehmann2015message}, 
a factor-graph approach is proposed for joint channel estimation, self-interference mitigation, 
and decoding, which is suitable for OFDM-based self-interference-limited transmissions.

\subsection{Adaptive Signal Optimization}
According to the domains of the decision variables, different ISAC signal parameter optimization algorithms are proposed.
Adaptive signal optimization methods are proposed for various
scenarios, including ISAC signal parameter optimization
and ISAC signal structure optimization. When optimizing
the parameters of ISAC signal, the objective functions and
constraints need to be designed according to the requirements
of scenarios, including sensing MI, DIR, PAPR, and transmit
power, etc. ISAC signal optimization faces the trade-off between
the performance of sensing and communication, which is
a challenge of ISAC signal optimization \cite{zilz2017statistical}. According to the decision variables, different ISAC signal
parameter optimization algorithms are proposed.

\subsubsection{ISAC signal parameter optimization}
The parameters of ISAC signal in space-time-frequency domains need to be optimized to satisfy the requirements of communication and sensing in various application scenarios.

\begin{itemize}
	\item[$\bullet$] Parameter optimization in space domain
\end{itemize}

The beamforming and precoding methods have been proposed for the ISAC signal optimization 
in the space domain. In \cite{tang2019analysis}, the impact of random communication signal on 
radar peak sidelobe level (PSL) is reduced by selecting the signal with high PSL to 
detect weak targets accurately. The transmit
waveform matrix is optimized for designing the radar beam pattern under the 
constraints of transmit power and signal-to-interference plus noise ratio (SINR) in \cite{liu2018mu}. 
The CRLB of sensing channel estimation is minimized under the constraint of SINR 
for each user \cite{liu2018toward}. In \cite{hua2021optimal}, 
a low-complexity minimum beam pattern gain maximization method is proposed to 
match the beam pattern and maximize the minimum weighted beam
pattern gain.

\begin{itemize}
	\item[$\bullet$] Parameter optimization in time-frequency domains
\end{itemize}

In terms of the ISAC signal optimization in time-frequency domains, existing works applied the sensing and communication MI and CRLB to optimize the ISAC signal.

Several literature take sensing MI as the objective function of the ISAC signal optimization.
For example, the sensing MI is maximized under the constraints of DIR and subcarrier
power ratio to improve the performance of ISAC systems in \cite{zhang2019mutual}.
The weighted sum of sensing MI and communication MI is maximized under the total
power constraint to obtain a closed-form solution for optimal power allocation of ISAC signals
\cite{yuan2020spatio, liu2017adaptive}.

Besides sensing and communication MI, other performance metrics are adopted as 
the objective functions. \cite{6494422} maximizes the detection probability of radar sensing 
under the constraints of the total transmit power and DIR. In point-to-point communication scenarios, 
the power of the ISAC signals is optimally allocated under the constraints of 
pulse compression sidelobe and communication data rate in \cite{keskin2021peak}. 
The proposed method in \cite{keskin2021peak} outperforms traditional windowing and 
waterfilling techniques under severe channel fading conditions. 
In \cite{zhu2021adaptive}, an adaptive ISAC signal optimization model with the maximum entropy, 
SNR and DIR are established under the
constraint of total power. To achieve a balanced trade-off between sensing and communication, 
the CRLB of sensing channel estimation is minimized to enhance the sensing accuracy and 
channel capacity under the constraint of total transmit
power \cite{liu2017multiobjective}.
To solve the optimization model, they propose
two signal optimization methods,
namely the weighted optimization ISAC signal design and
the Pareto optimal ISAC signal design method.

\subsubsection{ISAC signal structure optimization}
\begin{figure}[!htbp]
	\centering
	\includegraphics[width=0.5\textwidth]{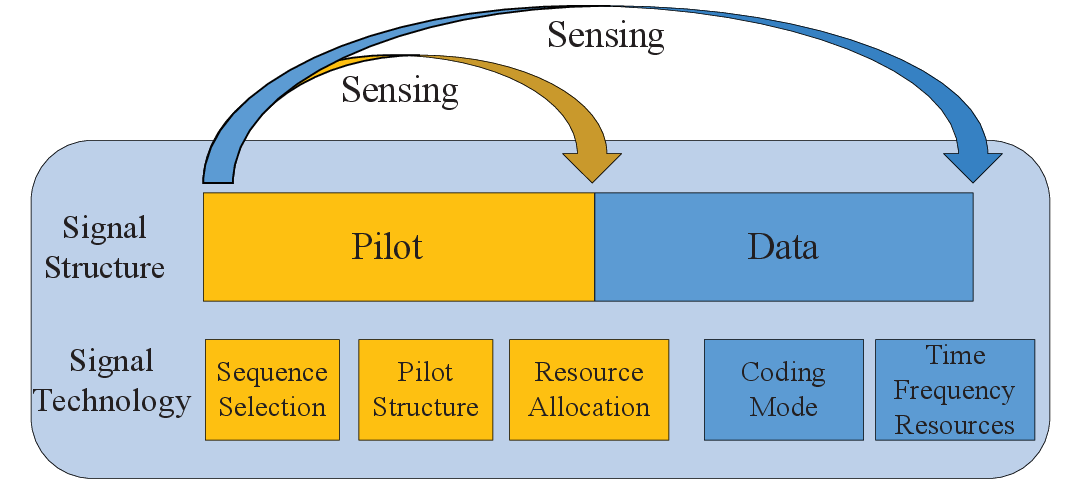}
	\caption{ISAC signal structure.}
	\label{Signal structure}
\end{figure}
The signal structure is divided into the pilot and data parts, 
as shown in Fig. \ref{Signal structure}. 
The pilot signal is known in the modulation domain and is normally inserted in the OFDM signal 
for synchronization and channel estimation \cite{mousaei2017comsens}. 
The pilot signal has the advantages
of high power, excellent sensing performance and strong anti-interference capability,
which has the potential to be applied in radar sensing \cite{wei20225g}.
The performance of the ISAC signal is improved by optimizing the pilot signal structure.
Pilot sequences with ideal autocorrelation characteristics are used to
improve the radar sensing. In addition, the structure and resource allocation of
the pilot also affect the performance of radar sensing.

In terms of sequence selection,
sequences with ideal autocorrelation characteristics and good cross-correlation characteristics 
should be selected. In \cite{kumari2017ieee},
the preamble composed of Golay sequence of 802.11ad is applied to radar sensing
because of its good autocorrelation properties. In \cite{liu2019integration},
the Zadoff-Chu sequence is used for radar processing because of its good orthogonal ability,
which reduces the impact of phase between adjacent pulses.
In \cite{wei20225g}, the performance of 5G reference signals,
including positioning reference signal (PRS), demodulation reference signal (DMRS),
channel state information-reference signal (CSI-RS) and synchronization signal (SS),
is verified and it is proved that PRS has advantages in radar sensing.
Further, the concept of sensing reference signal is proposed by exploiting the existing
reference signal in radar sensing, which is specifically suitable to 5G-A.

\begin{figure}[!htbp]
    \centering
    \includegraphics[width=0.5\textwidth]{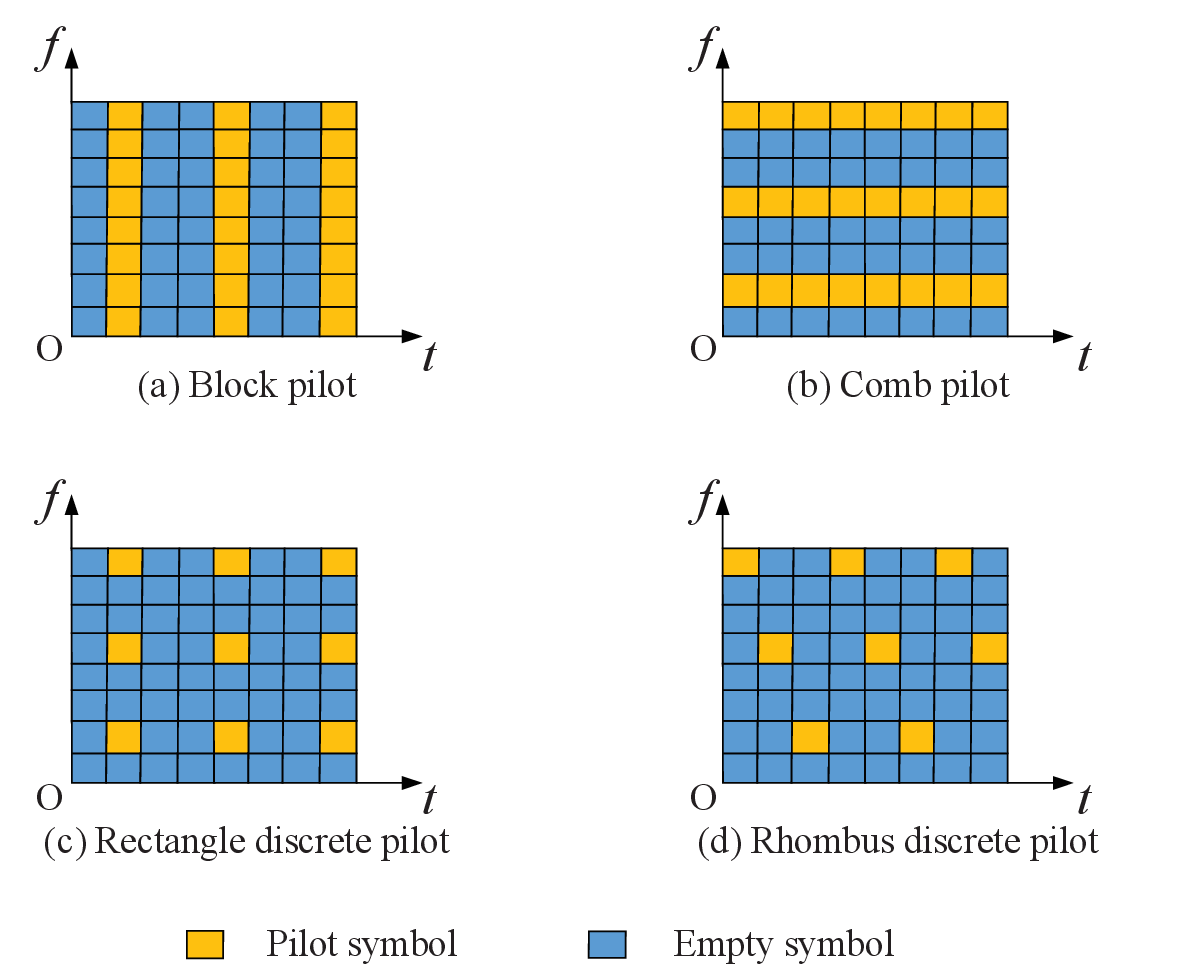}
    \caption{Main pilot structures.}
    \label{Common pilot structures}
\end{figure}

In terms of the pilot structure, due to the 2D time-frequency properties of OFDM signals, 
the pilot signals are inserted in the 2D resource blocks. There are three common pilot structures 
as shown in Fig. \ref{Common pilot structures}. Block pilots are distributed continuously in 
frequency domain and discretely in time domain, which are suitable for 
frequency-selective fading channels. The comb pilots are continuously distributed in
time domain and discretely in frequency domain.
Comb pilots are suitable for time-selective fading channels.
Discrete pilots are discretely inserted in both the time and
frequency domains, which saves time-frequency resources and
significantly improves spectrum efficiency \cite{ozkaptan2018ofdm,bao2020superimposed}.

The comb pilot structure is adopted in \cite{ozkaptan2018ofdm}, 
where the pilot sequence is Barker code and the radar signal processing method combining matched 
filter detection and 2D Fourier transform is applied. 
The distance resolution of 0.1 m and velocity resolution of 0.58 m/s are realized.
In \cite{bao2020superimposed}, an OFDM-MIMO ISAC system using a superimposed pilot is proposed. 
Compared with comb pilot and block pilot, 
the superimposed pilot has the advantages of comb pilots in velocity estimation and 
block pilots in distance estimation, which is suitable for 
frequency-selective and fast fading channels.

In terms of resource allocation,
the subcarrier allocation between the data and pilot is optimized to maximize
the sensing accuracy
and channel capacity \cite{ozkaptan2018ofdm}.
In \cite{wang2020multi}, the dynamic change of the number of pilot subcarriers is used for
radar detection at different distances, which realizes short-range radar (SRR),
medium-range radar (MRR) and long-range radar (LRR).
A full-duplex ISAC system is proposed in \cite{soury2017optimal},
where the pilot overhead is used for channel estimation and radar sensing,
and the sum data rate is
optimized according to the pilot overhead and power.

In addition, as shown in Fig. \ref{Signal structure}, the autocorrelation of the data
part is improved by utilizing phase coding to enhance the sensing performance \cite{tigrek2007golay}. Radar sensing with both pilot and data achieving high sensing performance is the future trend in ISAC signal design and optimization.

\section{Future Trends}

\begin{figure}[!htbp]
	\centering
	\includegraphics[width=0.5\textwidth]{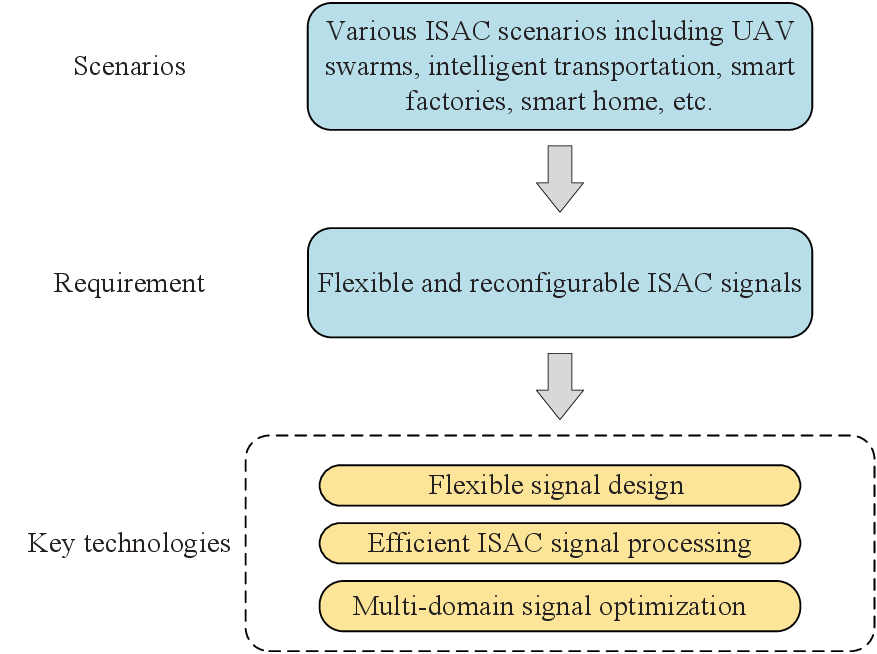}
	\caption{Future research trends of ISAC signals.}
	\label{futurework}
\end{figure}

The application scenarios of ISAC systems are various. 
As shown in Fig. \ref{Rich ISAC scenarios}, typical scenarios include UAV swarm, intelligent
transportation, smart factory, smart home, etc. Therefore,
the ISAC signals need to be flexible and reconfigurable,
adapting to diverse scenarios. As shown in Fig. \ref{futurework}, this
section provides the future research trends of ISAC signals for
various application scenarios, including signal design, signal
processing and signal optimization, to realize the goal of
flexible and reconfigurable ISAC signals.

\subsection{ISAC Signal Design}

ISAC signal design mainly includes waveform design, frame structure design,
and transmission mode design. Academics continue to propose new waveforms for 5G-A and 6G.
It is crucial to select the appropriate waveform for ISAC through performance evaluation. It is necessary to optimize the frame structure to improve the efficiency of communication and sensing.
In terms of transmission mode, the duplex method is still an open problem.

\subsubsection{Waveform design}
OFDM is the mainstream modulation technology in mobile communication systems. 
However, various waveforms, such as FBMC, GFDM, and OTFS, 
are proposed for the next-generation mobile communication systems, which are the further trends of modulation technology.
Considering the high DIR and sensing accuracy of the signals on the THz spectrum bands, 
the ISAC signals on THz bands have great advantages in improving the performance of sensing and communication. 
It is noted that the candidate ISAC signals on THz bands 
include DFT-s-OFDM and single carrier-frequency domain equalization (SC-FDE), etc.

As part of the Internet of things (IoT) applications, it is expected that ISAC technology will be required not only on the base station (BS) side for sensing the surrounding environment, such as intelligent transportation and other application scenarios, but also on the terminal side in various IoT scenarios, such as smart home.

Specific requirements need to be considered in the modulation technology. For example, low sidelobe, high reliability\cite{zhang2022integrated} and multi-user interference\cite{wang2021joint}. In addition, due to the different requirements of communication and sensing, it is necessary to balance communication and sensing when designing the ISAC signal.

\subsubsection{Frame structure}
Existing ISAC signals generally adopt the frame structure specified by 
the standards such as 3GPP 5G NR and IEEE 802.11. In complex scenarios, the fixed frame structure cannot perfectly adapt to differentiated requirements. Hence, the frame structure design of ISAC signals should be flexible and reconfigurable to adapt to various scenarios. The existing ISAC
signals generally apply pilots to realize the sensing function,
which will not affect the communication function. When more
resources are allocated to the pilot, the sensing performance
will be improved, and the communication performance will be
degraded. There are some studies on the flexible design of the
ISAC frame structure. In \cite{ozkaptan2018ofdm}, the pilot and data subcarriers
are optimally allocated to improve the performance of sensing
and communication. The pilot-based ISAC signal in \cite{babenko2019efficient}
flexibly allocate the pilots to achieve long-distance sensing 
\cite{babenko2019efficient}. In the future, joint optimization of the parameters in space-time-frequency domains adapts to various scenarios and realize flexible
and reconfigurable ISAC signals.

\subsubsection{Transmission mode}
The transmission modes of communication include full duplex and half duplex.
The majority of ISAC systems adopt half duplex mode.
\cite{xiao2022waveform}
propose a full duplex (FD) ISAC scheme,
which uses the waiting time of traditional pulse radar to transmit communication signals.
Compared with the traditional ISAC signal,
this scheme improves the DIR and solves the problem of blind detection.
However, the transmission mode suitable to ISAC is still one of the future research trends.

\subsection{ISAC Signal Processing}
ISAC signal processing includes single-BS sensing  algorithms
and multi-BS cooperative sensing algorithms.
High sensing accuracy and low computational complexity are the
two most critical performance metrics pursued by ISAC signal processing algorithms.
ISAC signal processing algorithm will be the trade-off between these two performance indices.
For new applications such as intelligent transportation and smart city,
multi-BS cooperative sensing is the
key technology to realize large-scale and high-accuracy sensing.
The design of a cooperative sensing algorithm is challenging.

\subsubsection{ISAC signal processing with high accuracy and low complexity}

There is a trade-off between the complexity and accuracy
when designing ISAC signal processing algorithms.
\cite{ma2022downlink} analyzes the sensing performance of the existing frame structure using different reference signals. However, since there are multiple reference signals,
the effective application of these reference signals is still an open problem.
In the future, the ISAC systems on the mmWave and THz bands will be common \cite{li2021integrated}.
Hence, the high-accurate ISAC signal processing algorithms on mmWave and
THz bands are the future research trends.

\subsubsection{Cooperative signal processing}
The networked mobile communication system is applied to realize cooperative sensing using multiple BSs, which can significantly improve the
sensing accuracy and range \cite{jiang2022improve}.
Correspondingly, the signal processing for cooperative sensing needs to be studied.
Most existing cooperative sensing methods are designed for pulse radar or coherent radar.
Take the coherent radar as an example, the signals transmitted by each radar are adjusted to the same frequency and phase when they reach the target,
which greatly improves the SNR of the echo signal \cite{yin2014wideband}.
Coherent radar requires high time and phase synchronization,
whose signal processing methods are unsuitable for mobile
communication systems with ISAC capability.
There are few existing studies on cooperative sensing methods for ISAC systems.
Hence, the signal processing for cooperative sensing and communication system
is one of the future trends.

\subsection{ISAC Signal Optimization}
The ISAC signal design mainly focuses on the research of modulation technology,
and there is no flexible and reconfigurable ISAC design combined with the frame structure.
Besides, there are no comprehensive signal optimization schemes
in the space-time-frequency domains.
The goal of spatial optimization is to design the ISAC beams into ideal directions
to improve the performance of sensing and communication.
Optimization in the time-frequency domains focuses on
the subcarrier and power allocation of
ISAC signals.
The ISAC signal optimization in the space-time-frequency domains
simultaneously optimizes the precoding matrix, subcarrier allocation, power allocation,
frame structure, etc., to obtain an ISAC signal
with approximate ideal performance.
In addition, there are limited studies on ISAC signal optimization for
multi-BS cooperative sensing.
Overall, under the differentiated requirements of communication and
sensing in diversified scenarios,
the ISAC signal optimization in space-time-frequency multi-domain
is the future trend. Besides, to extend the sensing range and improve the sensing accuracy,
further in-depth research on ISAC signal optimization for cooperative sensing
is required.

\section{Conclusion}
ISAC has been regarded as one of the potential key technologies of future mobile communication systems, with ISAC signal as the fundamental technology.
In this article, we review ISAC signals from the perspective of 5G, 5G-A and 6G mobile communication systems from three aspects, namely signal design, signal processing, and signal optimization. In the future, facing the various scenarios in the era of 5G-A and 6G, flexible and reconfigurable ISAC signals need to be designed.
This article comprehensively reviews the ISAC signals, which may provide a guideline for researchers in the area of ISAC
technology.

\bibliographystyle{IEEEtran}
\bibliography{reference}

\ifCLASSOPTIONcaptionsoff
  \newpage
\fi

\vspace{0.5 mm}
\begin{IEEEbiography}[{\includegraphics[width=1.1in,height=1.25in,clip,keepaspectratio]{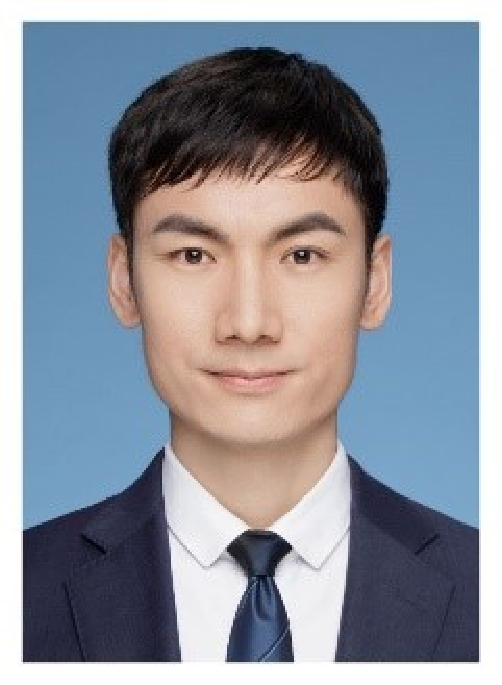}}]{Zhiqing Wei (Member, IEEE)} received the B.E. and Ph.D. degrees from the Beijing University of Posts and Telecommunications (BUPT), Beijing, China, in 2010 and 2015, respectively. He is an Associate Professor with BUPT. He has authored one book, three book chapters, and more than 50 papers. His research interest is the performance analysis and optimization of intelligent machine networks. He was granted the Exemplary Reviewer of IEEE WIRELESS COMMUNICATIONS LETTERS in 2017, the Best Paper Award of WCSP 2018. He was the Registration Co-Chair of IEEE/CIC ICCC 2018, the publication Co-Chair of IEEE/CIC ICCC 2019 and IEEE/CIC ICCC 2020.
\end{IEEEbiography}

\vspace{0.5 mm}
\begin{IEEEbiography}[{\includegraphics[width=1.1in,height=1.25in,clip,keepaspectratio]{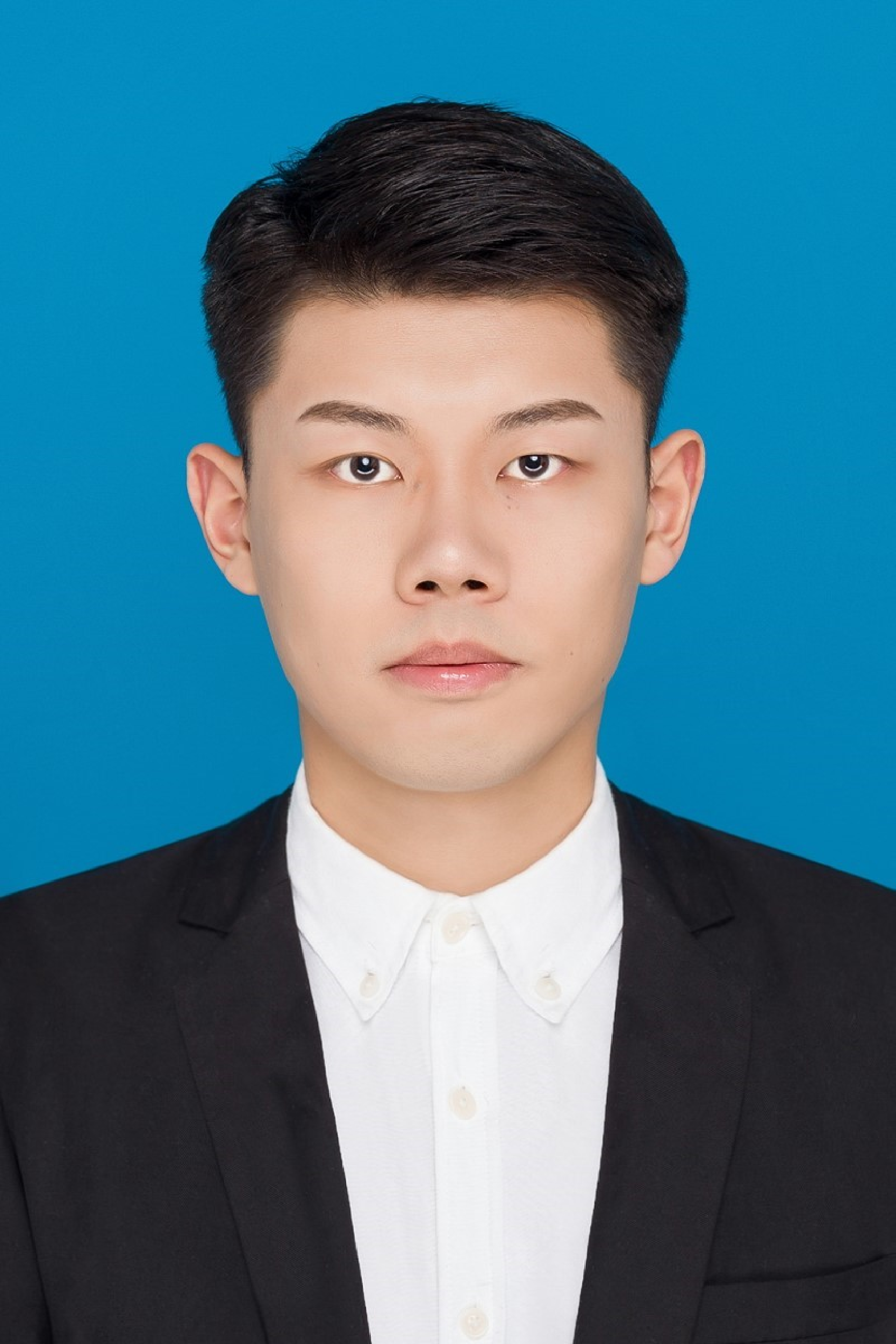}}]{Hanyang Qu (Student Member, IEEE)} received the B.S. degree in School of Electronic Engineering and Optoelectronic Technology, Nanjing University of Science and Technology (NJUST) in 2020. He is currently pursuing his master's degree with Beijing University of Posts and Telecommunication (BUPT). His research interests include integrated sensing and communication and high accuracy sensing.
\end{IEEEbiography}

\vspace{0.5 mm}
\begin{IEEEbiography}[{\includegraphics[width=1in,height=1.15in,clip,keepaspectratio]{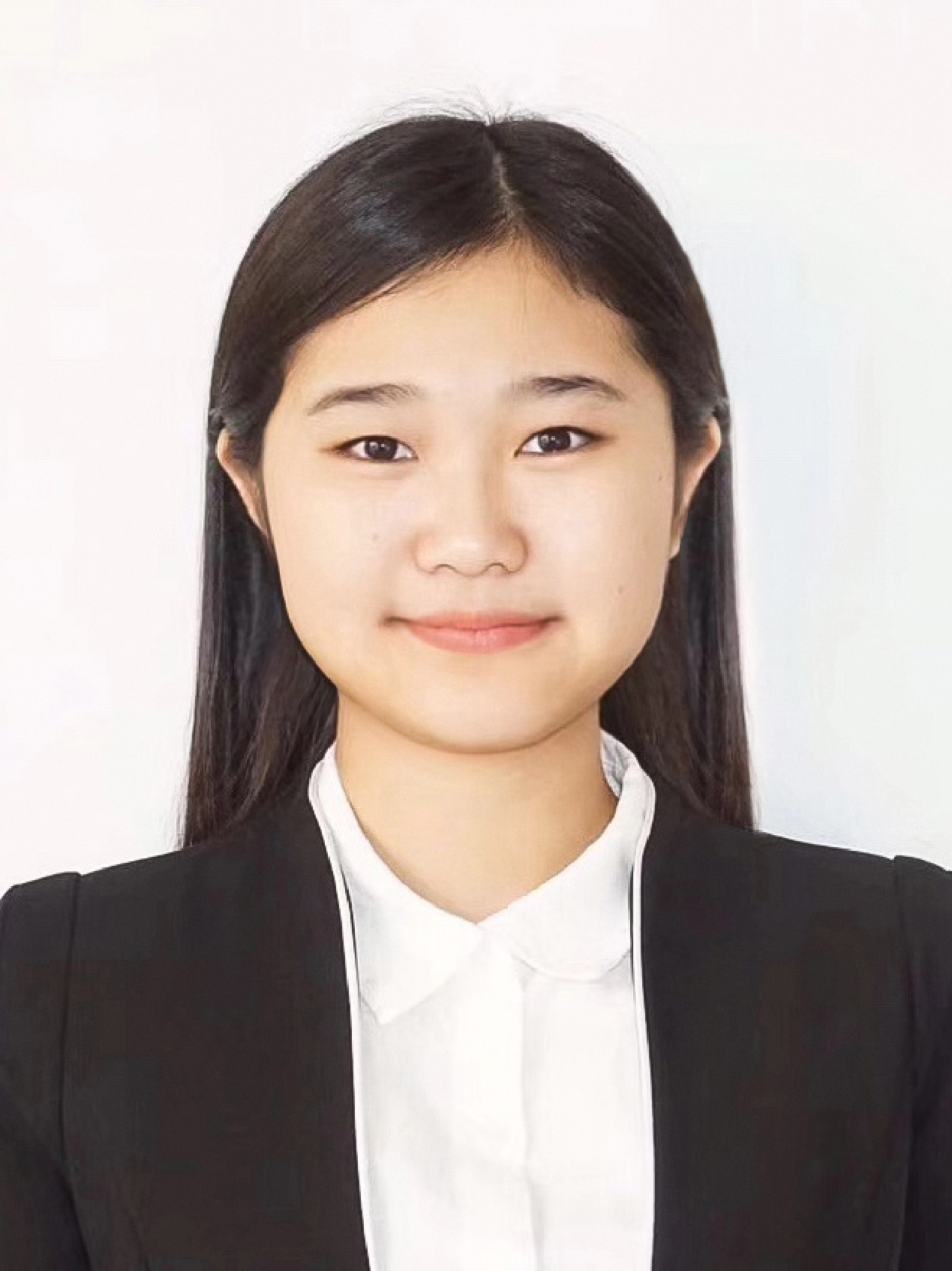}}]{Yuan Wang}
	received her B.E. degree from University of Science \& Technology Beijing, USTB, Beijing, China. She is currently pursuing the M.E. degree with the School of Information and Communication Engineering, Beijing University of Posts and Telecommunications (BUPT), Beijing, China. Her research interests include integrated sensing and communication.
\end{IEEEbiography}

\vspace{0.5 mm}
\begin{IEEEbiography}[{\includegraphics[width=0.95in,height=1.1in,clip,keepaspectratio]{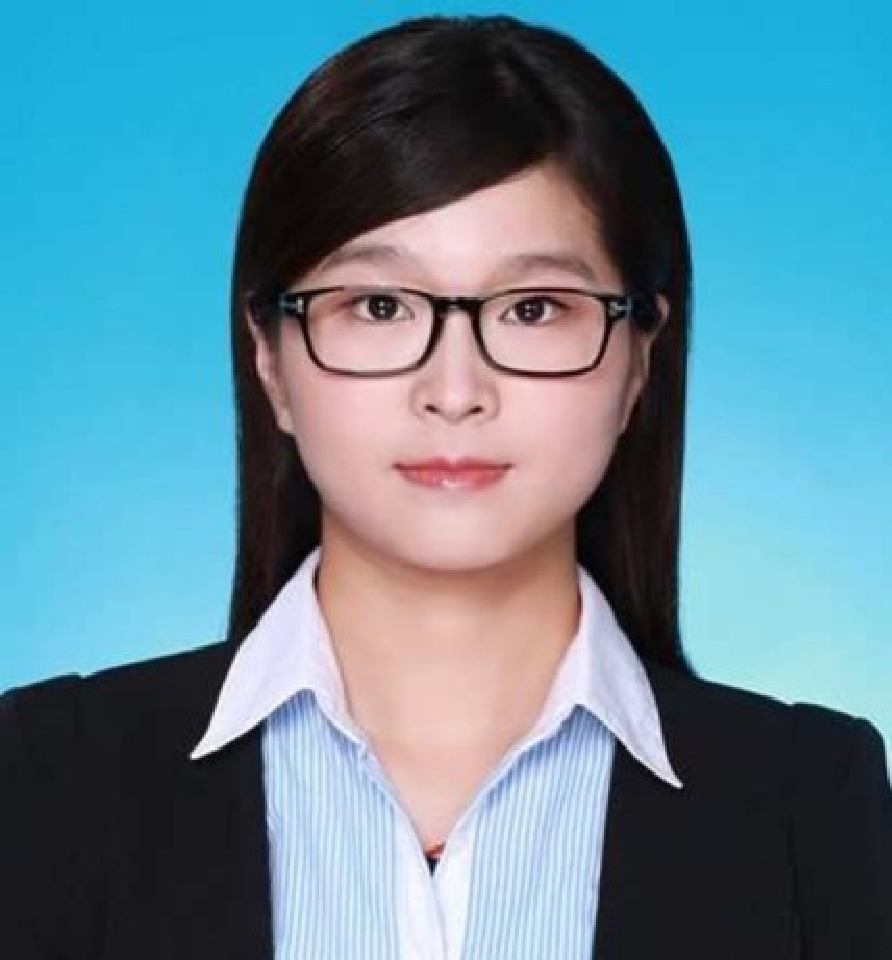}}]{Xin Yuan (Member, IEEE)} received the B.E. degree from Taiyuan University of Technology, Shanxi, China, in 2013, and the dual Ph.D. degree from Beijing University of Posts and Telecommunications (BUPT), Beijing, China, and the University of Technology Sydney (UTS), Sydney, Australia, in 2019 and 2020, respectively. She is currently a Research Scientist at CSIRO, Sydney, NSW, Australia. Her research interests include machine learning and optimization, and their applications to UAV networks and intelligent systems.
\end{IEEEbiography}

\vspace{0.5 mm}
\begin{IEEEbiography}[{\includegraphics[width=1.25in,height=1.4in,clip,keepaspectratio]{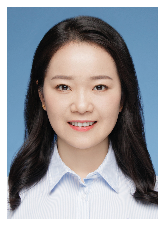}}]{Huici Wu (Member, IEEE)} received the Ph.D degree from Beijing University of Posts and Telecommunications (BUPT), Beijing, China, in 2018. From 2016 to 2017, she visited the Broadband Communications Research (BBCR) Group, University of Waterloo, Waterloo, ON, Canada. She is now an Associate Professor at BUPT. Her research interests are in the area of wireless communications and networks, with current emphasis on collaborative air-to-ground communication and wireless access security.
\end{IEEEbiography}

\vspace{0.5 mm}
\begin{IEEEbiography}[{\includegraphics[width=1.1in,height=1.25in,clip,keepaspectratio]{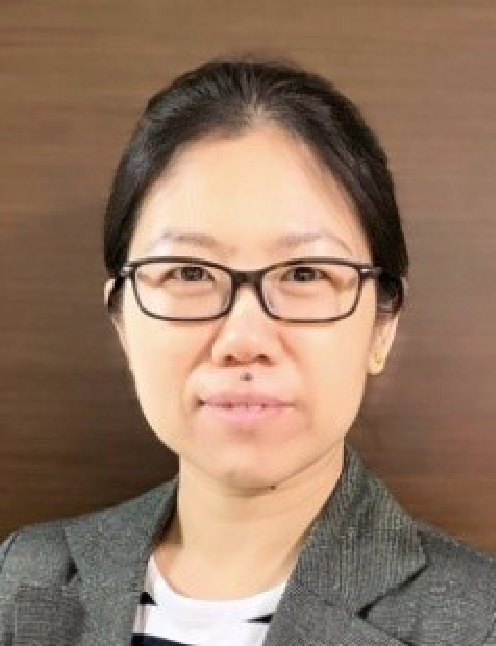}}]{Ying Du} is currently pursuing the Ph.D. degree with University of Science and Technology of China. She is the professor-level senior engineer of the China Academy of Information and Communications Technology (CAICT). She is vice chair of Standards and International Cooperation Working Group of IMT-2030(6G) Promotion Group. She has contributed to the research, evaluation and international standardization of IEEE 802.16, 4G, 5G and 6G communication systems. She has authored or co-authored more than 30 research papers or articles, and over 40 patents in this area. She has hosted three National Major Projects on mobile broadband system. Currently, she focuses on technology research, standardization, and verficaiton for 5G-Advanced and 6G systems.
\end{IEEEbiography}

\vspace{0.5 mm}
\begin{IEEEbiography}[{\includegraphics[width=1.1in,height=1.25in,clip,keepaspectratio]{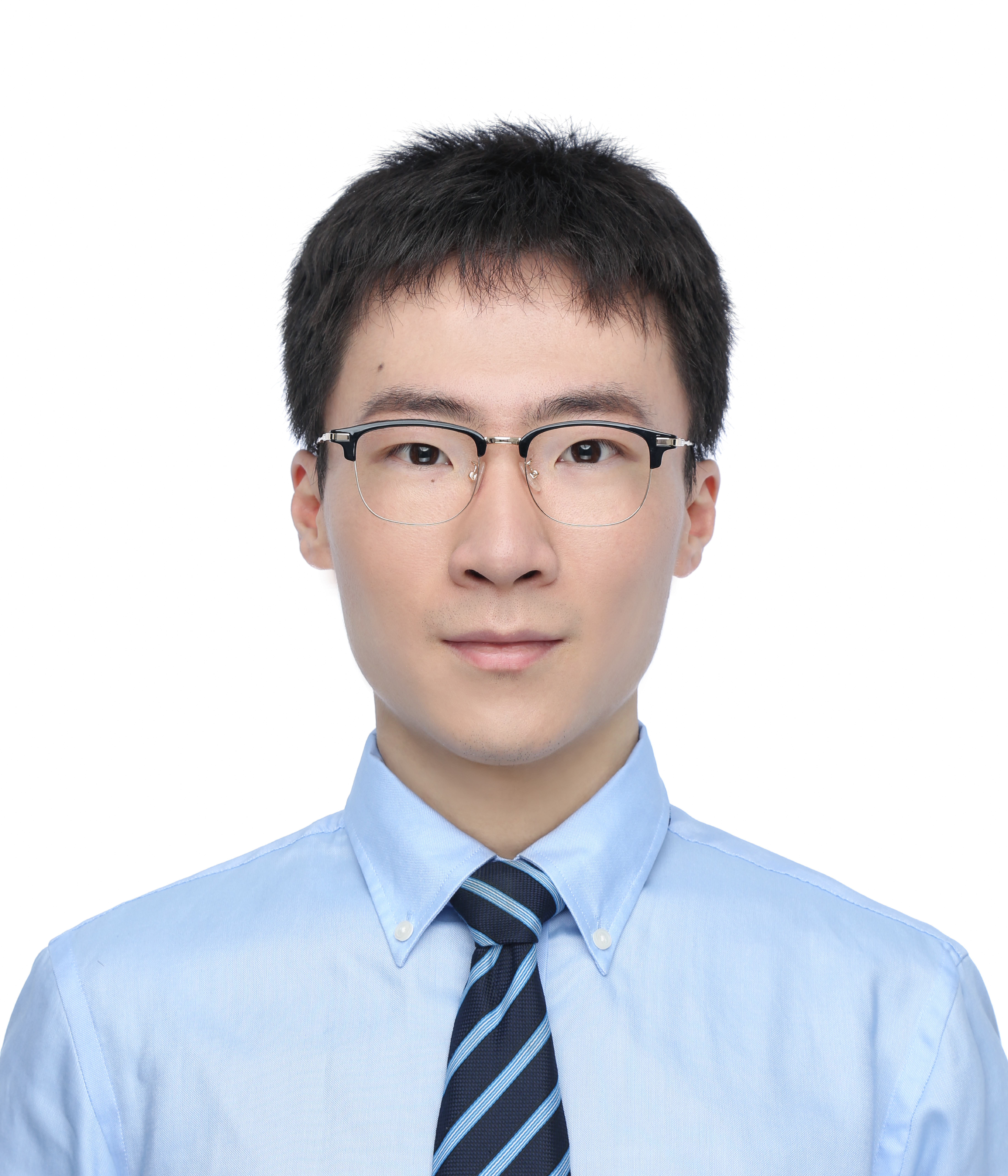}}]{Kaifeng Han (Member, IEEE)} is a senior engineer in the China Academy of Information and Communications Technology (CAICT). Before that, he obtained his Ph.D. degree from the University of Hong Kong in 2019, and the B.Eng. (first-class hons.) from the Beijing University of Posts and Telecommunications and Queen Mary University of London in 2015, all in electrical engineering. His research interests focus on integrated sensing and communications, wireless AI for 6G. He is funded by China Association for Science and Technology (CAST) Young Talent Support Program. He received one best paper award and published 40+ research papers in international conferences and journals.
\end{IEEEbiography}

\vspace{0.5 mm}
\begin{IEEEbiography}[{\includegraphics[width=1.1in,height=1.25in,clip,keepaspectratio]{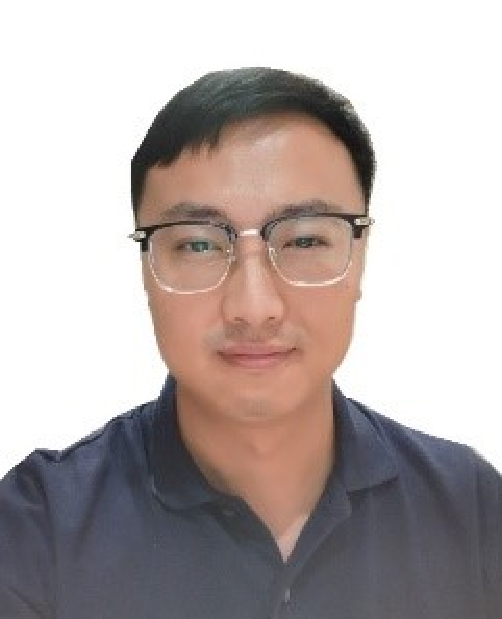}}]{Ning Zhang (Senior Member, IEEE)}  received the Ph.D degree in Electrical and Computer Engineering from University of Waterloo, Canada, in 2015. After that, he was a postdoc research fellow at University of Waterloo and University of Toronto, respectively. Since 2020, he has been an Associate Professor in the Department of Electrical and Computer Engineering at University of Windsor, Canada. His research interests include connected vehicles, mobile edge computing, wireless networking, and security. He is a Highly Cited Researcher (Web of Science). He serves/served as an Associate Editor of IEEE Transactions on Mobile Computing, IEEE Internet of Things Journal, IEEE Transactions on Cognitive Communications and Networking, and IEEE Systems Journal. He also serves/served as a TPC chair for IEEE VTC 2021 and IEEE SAGC 2020, a general chair for IEEE SAGC 2021, a chair for track of several international conferences and workshops including IEEE ICC, VTC, INFOCOM Workshop, and Mobicom Workshop. He received 8 Best Paper Awards from conferences and journals, such as IEEE Globecom, IEEE ICC, IEEE ICCC, IEEE WCSP, and Journal of Communications and Information Networks. He also received IEEE TCSVC Rising Star Award and IEEE ComSoc Young Professionals Outstanding Nominee Award. He has been an IEEE senior member since 2018.
\end{IEEEbiography}

\vspace{0.5 mm}
\begin{IEEEbiography}[{\includegraphics[width=1.1in,height=1.25in,clip,keepaspectratio]{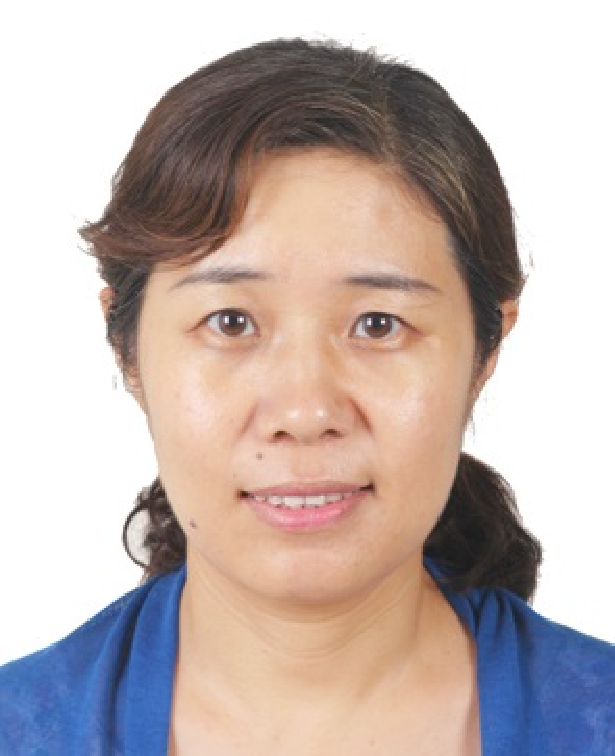}}]{Zhiyong Feng} (M'08-SM'15) received her B.E., M.E., and Ph.D. degrees from Beijing University of Posts and Telecommunications (BUPT), Beijing, China. She is a professor at BUPT, and the director of the Key Laboratory of the Universal Wireless Communications, Ministry of Education, P.R.China. She is a senior member of IEEE, vice chair of the Information and Communication Test Committee of the Chinese Institute of Communications (CIC). Currently, she is serving as Associate Editors-in-Chief for China Communications, and she is a technological advisor for international forum on NGMN. Her main research interests include wireless network architecture design and radio resource management in 5th generation mobile networks (5G), spectrum sensing and dynamic spectrum management in cognitive wireless networks, and universal signal detection and identification.
\end{IEEEbiography}

\end{document}